# **M**etamodel **Qua**lity **R**equirements and **E**valuation (MQuaRE)

Taciana N. Kudo[1,2]

Renato F. Bulcão-Neto[1]

Auri M. R. Vincenzi[2]

[1]Instituto de Informática, UFG, Goiânia-GO, Brazil

{taciana,rbulcao}@ufg.br

[2]Departamento de Computação, UFScar, São Carlos-SP, Brazil

auri@dc.ufscar.br

V 2.0

September, 2020

# Revision History

| Date | Version | Description | Author |
|---|---|---|---|
| 19th August | 1.0 | First version of MQuaRE documentation | Taciana Novo Kudo<br>Renato F. Bulcão-Neto<br>Auri M. R. Vincenzi |
| 1st September | 2.0 | Changes in the MQuaRE annexes | Taciana Novo Kudo<br>Renato F. Bulcão-Neto |
|  |  |  |  |
|  |  |  |  |

# Summary

# List of Figures

# LIST OF TABLES

# 1. INTRODUCTION

## 1.1. Scope

Models are the primary artifacts of model-driven software engineering (MDSD) [1], and a terminal model is a representation that conforms to a given software metamodel [2, 3]. As the quality of a software metamodel directly impacts the quality of terminal models, software metamodel quality is an essential aspect of MDSD.

However, the literature reports a few proposals for metamodel quality evaluation, but most lack a general solution for the quality issue. Some efforts focus on quality measures [4], a quality evaluation model [5], or a quality evaluation model with structural measures borrowed from OO design [6-8]. Thus, we support there is a need for a more thorough solution for metamodel quality evaluation, with potential benefits to MDSD in general.

This document describes a metamodel quality evaluation framework called MQuaRE (Metamodel Quality Requirements and Evaluation). MQuaRE is an integrated framework composed of metamodel quality requirements, a metamodel quality model, metamodel quality measures, and an evaluation process, with a great contribution of the ISO/IEC 25000 series [9] for software product quality evaluation.

## 1.2 Organization of MQuaRE

The Metamodel Quality Requirements and Evaluation (MQuaRE) is a complete proposal to guide software metamodels quality evaluation. The complete view of MQuaRE is presented in the Figure 1 and includes:

- **The MQuaRE's quality requirements**: MQuaRe offer 19 (nineteen) requirements that an metamodel user and an evaluator can use to consider the metamodel quality.
- **The MQuaRE's quality model:** MQuaRe offer 5 (five) characteristic and 10 (ten) sub-characteristics of metamodel quality that drive the documentation of metamodel quality requirements.
- **The MQuaRE's quality measures:** MQuaRe offer 23 (twenty three) measures to quantify the quality characteristics and sub-characteristics applying predefined measurement functions.
- **The MQuaRE's process**: MQuaRe describes a process with 5 (five) activities that define how quality model and requirements must be used in an evaluation activities, and when the measures will be applied to calculate the quality values. Besides that, the process defines the tasks input and output artifacts , and users' roles.

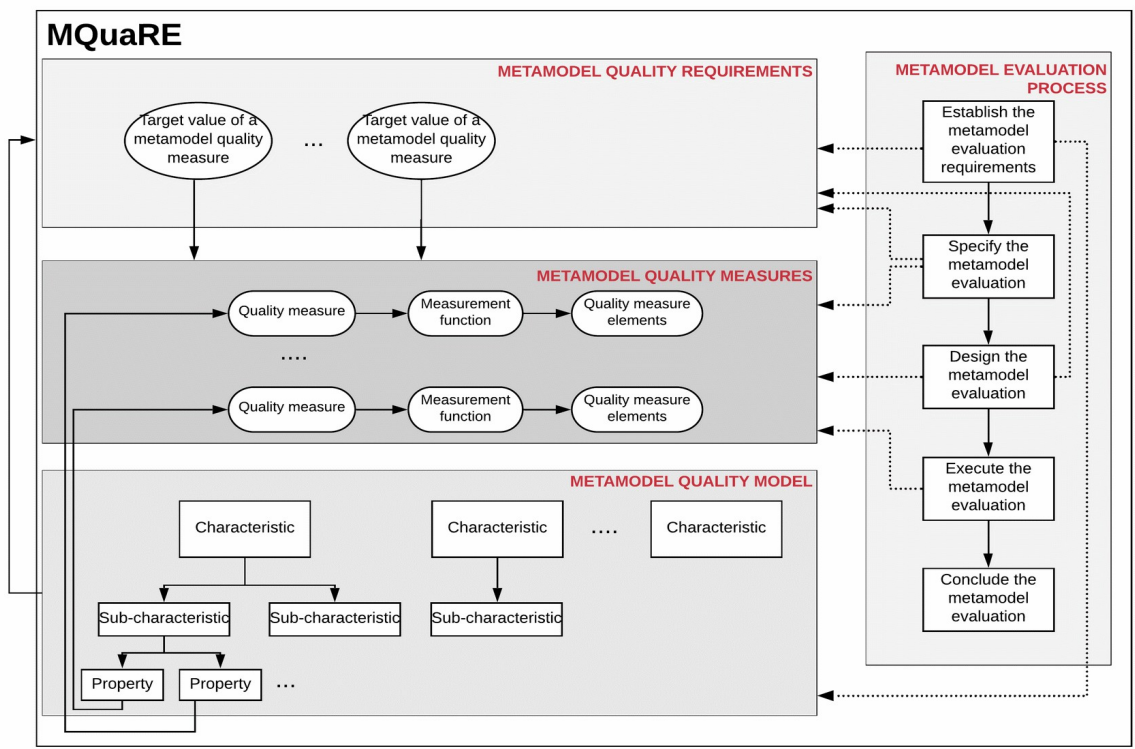

**Figure 1: An overview of MQuaRE: evaluation process and quality requirements, measures, and model.**

## 1.3. TERMS AND DEFINITIONS

For the purposes of this document, the following terms and definitions apply.

. *concept:* foundation elements of a metamodel (metamodel elements), e.g., a concept may be a class, relationship, or attribute in a UML metamodel. As we propose MQuARE as a generic-purpose framework, concepts may vary according to the specification language of the metamodel under evaluation.

. *evaluator:* person that performs a metamodel quality evaluation. [10]

. *evaluation requester:* person that requests a metamodel quality evaluation. [10]

. *evaluation tool[1]:* instrument that can be used during a metamodel quality evaluation to collect data, to perform interpretation of data or to automate part of the evaluation. [10]

. *measure[2]:* variable to which a value is assigned as the result of measurement. [10]

. *measurement:* set of operations having the object of determining a value to be assigned to a measure. [11]

. *measurement function:* algorithm or calculation performed to combined the measure elements. [11]

. *metamodel element:* any things that are part of a modelling, such as the attributes, operations, relations, and semantics of a class.

. *metamodel quality:* degree to which the metamodel elements satisfy needs when used under specified conditions.

. *metamodel quality evaluation:* systematic examination of the extent to which a metamodel is capable of satisfying stated needs.

. *metamodel specification:* includes all types of metamodel specification, including requirements specification, design specification, user documentation, or all of these.

. *quality measure:* measure that is defined as a measurement function of two or more values of quality measure elements. [11]

. *quality measure element:* measure defined in terms of an attribute and the measurement method for quantifying it, including optionally the transformation by a mathematical function. [11]

. *quality property:* measurable component of quality. [12]

. *user documentation:* describes how a metamodel can be used and contemplates the usage scenarios. A user documentation can be a user manual, a tutorial, a wizard, a "how to" lists or all of these.

. *usage scenario:* a real-world example of how one or more stakeholders or organizations interact with the metamodel. It describes the steps, events, and/or actions which need to occur for instantiating the metamodel.

## 2. METAMODEL QUALITY REQUIREMENTS

Quality requirements specification plays a crucial role in the metamodel evaluation process. Should quality requirements are not stated clearly, a same metamodel may be interpreted and evaluated variously by different people. As a result, one achieves an inconsistent metamodel evaluation.

Metamodel requirements address both the inherent and the assigned property requirements. The inherent property requirements include functional requirements and quality requirements. Functional requirements include the domain-specific requirements. Quality requirements may also imply architectural and structural requirements, and they are defined accordingly the characteristics of compliance, conceptual suitability, usability, maintainability, and portability. The assigned property requirements are composed by managerial requirements, including requirements for version control, delivery deadlines, copyright, to name a few. Figure 2 provides a categorization of metamodel requirements. Metamodel quality requirements (MQR) may comprise multiples aspects of a metamodel, e.g., whether it is easy to use and maintain or compliant to specific standards, if applicable.

---

[1] For instance, CASE tools to create instances of a metamodel, checklists to collect inspection data, or spreadsheets to produce synthesis of measures.
[2] The term "measure" refers collectively to base measures, derived measures, and indicators.

| Metamodel inherent property requirements | Functional requirements |
|---|---|
| | **Metamodel quality requirements**: compliance, conceptual suitability, usability, maintainability and portability |
| Metamodel assigned property requirements | Development process requirements |

**Figura 2: Metamodel requirements categorization**

## 2.1 Pre-conditions for Metamodel Quality Requirements

A quality model drives the documentation of MQRs. Despite that, we recommend the following pre-conditions for MQR:

- MQR shall be uniquely identified and following the objective of the metamodel evaluation;
- MQR shall be associated with quality sub-characteristics, as defined in the MQuaRE's quality model;
- MQR shall be specified in terms of a quality measure and a target value, which is the acceptable value for fulfilling a particular MQR;
- An acceptable tolerance value for the target value of a particular MQR shall be documented;
- Specific concepts and terms used in the metamodel should be used to avoid misunderstandings of the MQR;
- MQR shall be validated and approved by an evaluation requester.

## 2.2 Metamodel Quality Requirements Verification

Defining the MQR is essential to avoid inconsistencies in the metamodel evaluation. Here is a list of recommendations to ensure the quality of MQR:

- MQR shall be verifiable, reviewed, and approved;
- Evaluation tools, techniques, or other resources (e.g., effort or time) required for verification shall be documented;
- Identified conflicts between MQR shall be documented;
- Identified conflicts between MQR or between MQR and metamodel concepts shall be documented;
- The stakeholders' identities shall be documented.

## 2.3 List of Metamodel Quality Requirements

MQuaRE provides 19 (nineteen) MQRs that meet the pre-conditions presented in Section 2.1 and can be reused by metamodel users and evaluators.

**MQR01** - The metamodel conceptual foundation must comply with widely-accepted and sound theories, regulations, standards, and conventions.

**MQR02** - The metamodel must cover the concepts found in its specifications.

**MQR03** - The metamodel must represent the concepts found in its specifications correctly.

**MQR04** - The metamodel must represent the concepts required for achieving specific usage objectives.

**MQR05** - The users must be able to recognize whether a metamodel is appropriate for their needs accordingly the usage scenarios described in the user documents.

> **MQR06** - The users must be able to recognize whether a metamodel is appropriate for their needs accordingly the demonstration features of metamodel concepts.
>
> **MQR07** - The users must be able to recognize whether a metamodel is appropriate for their needs accordingly the evident concepts to the user in the metamodel specifications.
>
> **MQR08** - The users must be able to recognize whether a metamodel contain concepts whose purpose is correctly understood without prior training.
>
> **MQR09** - The users must be able to recognize whether a metamodel is appropriate for their needs accordingly the metamodel user documentation.
>
> **MQR10** - The metamodel must be composed of discrete concepts such that a change of one concept has minimal impact on other concepts.
>
> **MQR11** - The metamodel must be composed of discrete concepts such that a creation of model elements does not enforce ordered modelling actions.
>
> **MQR12** - The metamodel must be able to be reused to modelling usage scenarios for different application domains.
>
> **MQR13** - The users must be able to recognize metamodel modifications acoordingly the changes documented in the metamodel specification during metamodel development life cycle.
>
> **MQR14** - The users must be able to recognize metamodel modifications acoordingly the change comments confirmed in review.
>
> **MQR15** - The metamodel must be reused modified without introducing inconsistencies or degrading metamodel quality.
>
> **MQR16** - The metamodel must be able to be adapted to modelling usage scenarios for different application domains.
>
> **MQR17** - The metamodel must be able to replace another specified metamodel for the same purpose in the same application domain, without introducing any additional learning or workaround.
>
> **MQR18** - The metamodel must be able to replace another specified metamodel for the same purpose in the same application domain, without degrading metamodel quality degree.
>
> **MQR19** - The metamodel must be able to replace another specified metamodel for the same purpose in the same application domain by using similar concepts of previous metamodel.

## 3. QUALITY MODEL

The quality of a metamodel is the degree to which it provides value to a modeling activity. These stated needs are represented in MQuaRE by a quality model that categorizes metamodel quality into characteristics, which in some cases, subdivide into sub-characteristics. This hierarchical decomposition provides a convenient breakdown of metamodel quality.

## 3.1. Structure used for the quality model

The measurable quality-related properties of a metamodel are called quality properties. It is necessary to identify a collection of properties that cover characteristics or sub-characteristics, obtain quality measures for each, and combine them to achieve a derived quality measure corresponding to the quality characteristic or sub-characteristic. Thus, the quality model allows the categorization of MQRs. Figure 3 shows the relationship between quality characteristics and sub-characteristics, and quality properties.

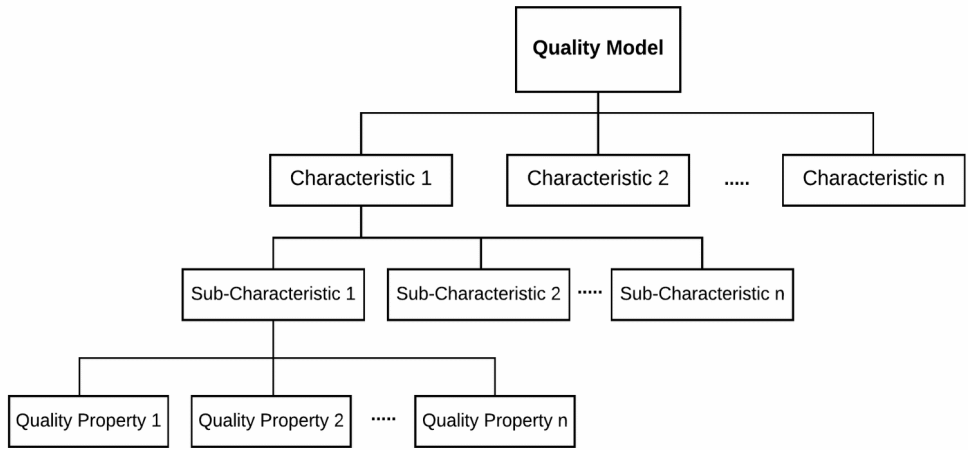

Figure 3: Commonly used structure for quality models.

## 3.2. The MQuaRE's quality model

The MQuaRE's quality model revises ISO/IEC 25010 [9], ISO/IEC 9126-1 [13], and related research [4-7], and incorporates quality characteristics and sub-characteristics with some amendments (see ANNEX A). Five characteristics form the MQuaRE's quality model, as depicted in Figure 4: Compliance, Conceptual Suitability, Usability, Maintainability, and Portability — further subdivided into sub-characteristics. These characteristics may work as a checklist for ensuring a comprehensive coverage of metamodel quality. Next, we describe the characteristics and sub-characteristics present in the MQuaRE's quality model.

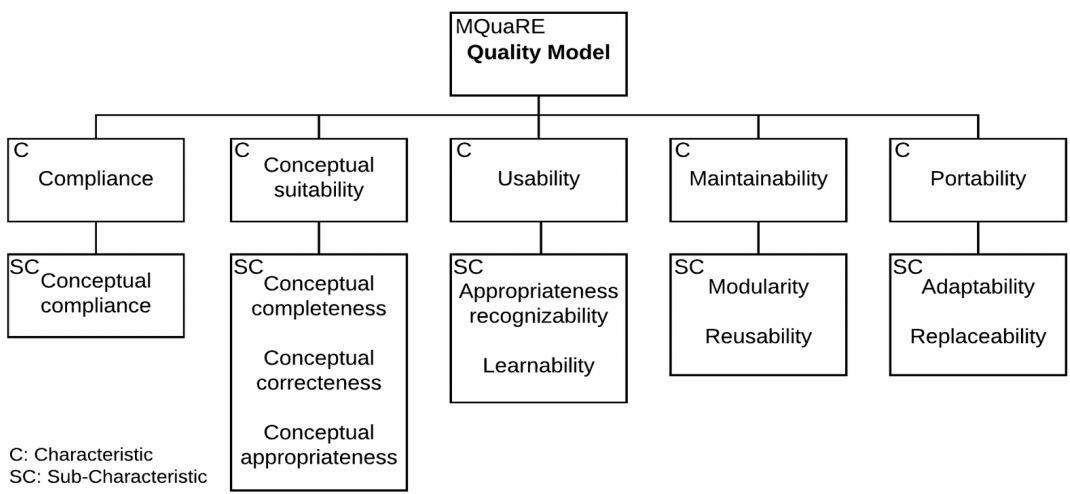

Figure 4: The MQuaRe's quality model with characteristics (C) and sub-characteristics (SC).

### 3.2.1 Compliance characteristic

The degree to which a metamodel must comply with items, such as widely accepted and sound theories, regulations, standards, and conventions. This characteristic includes the following sub-characteristic:

- **Conceptual compliance sub-characteristic**: the degree to which the conceptual foundation of a metamodel complies with widely accepted and sound theories, regulations, standards, and conventions.

### 3.2.2 Conceptual suitability characteristic

The degree to which a metamodel satisfies requirements when used under specified conditions. This characteristic includes the sub-characteristics below:

- **Conceptual completeness sub-characteristic:** the degree to which the set of metamodel concepts covers all the specified requirements.
- **Conceptual correctness sub-characteristic:** the degree to which the metamodel provides the correct modeling results with the needed degree of precision.
- **Conceptual appropriateness sub-characteristic:** the degree to which the metamodel facilitates the accomplishment of modeling tasks, and for determining their adequacy for performing these tasks.

### 3.2.3 Usability characteristic

The degree to which a metamodel can be used to achieve specific goals in a specified application domain. This characteristic includes the appropriateness recognizability and learnability sub-characteristics.

- **Appropriateness recognizability sub-characteristic:** the degree to which users can recognize whether a metamodel is appropriate for their needs or not.
- **Learnability sub-characteristic:** the degree to which a metamodel can be used by specified users to achieve specified learning goals in a given context of use.

### 3.2.4 Maintainability characteristic

The degree of effectiveness and efficiency with which a metamodel can be modified by the intended maintainers. This characteristic includes modularity, reusability, and modifiability sub-characteristics.

- **Modularity sub-characteristic:** the degree to which a metamodel is composed of discrete concepts such that a change of one concept has minimal impact on other concepts.
- **Reusability sub-characteristic:** the degree to which usage scenarios can be used in more than one metamodel.
- **Modifiability sub-characteristic:** the degree to which a metamodel can be effectively and efficiently modified without introducing inconsistencies or degrading existing metamodel quality.

### 3.2.5 Portability characteristic

The degree of effectiveness and efficiency with which a metamodel can be transferred from one application domain to another. This characteristic includes adaptability and replaceability sub-characteristics.

- **Adaptability sub-characteristic:** the degree to which a metamodel can effectively and efficiently be adapted for different application domains.
- **Replaceability sub-characteristic:** the degree to which a metamodel can replace another specified metamodel for the same purpose in the same application domain.

Therefore, the MQuaRE provides 5 quality characteristics and 10 ready-to-use sub-characteristics. Moreover, the quality model is flexible to be adapted to specific metamodels contexts, and new characteristics and sub-characteristics can be included.

## 4. METAMODEL QUALITY MEASURES

The quality characteristics and sub-characteristics can be quantified by applying measurement functions. A measurement function is a formula used to combine quality measure elements. The result of applying a measurement function is called a quality measure. In this way, quality measures are quantifications of the quality characteristics and sub-characteristics. Figure 5 illustrates the relationships between the quality model's components and the quality measures' elements.

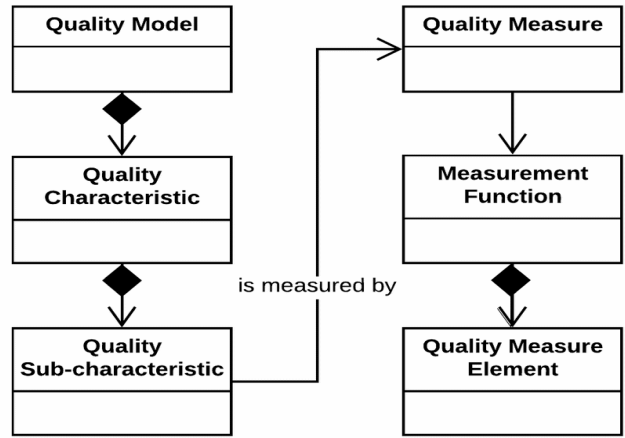

Figure 5: Relationship between quality model and measures.

## 4.1. The format used for quality measures documentation

Every quality measure is described by an identification code, the measure name, a description of the information provided by the measure, and a measurement function. The following is the definition of each data field of the quality measure documentation format in MQuaRE:

- ID: quality measure identification code consisting of two parts:
  - an abbreviated alphabetic code with the initial letter in uppercase of the quality characteristic followed by two letters representing the sub-characteristic. For instance, the ID UAp is for the quality measures of the Appropriateness Recognizability sub-characteristic of the Usability characteristic;
  - an ordinal number of the sequential order within a quality sub-characteristic. For instance, the ID UAp-2 means the second quality measure (i.e., demonstration coverage) of the Appropriateness Recognizability sub-characteristic of the Usability characteristic.
- Name: denomination used to refer a quality measure.
- Description: the information about the quality measure.
- Measurement function: showing how the *quality measures elements* are combined to produce the quality measure.

## 4.2. List of Metamodel Quality Measures

The current version of MQuaRE includes 23 (twenty-three) metamodel quality measures bound to the quality model as follows: Compliance (2), Conceptual Suitability (4), Usability (5), Maintainability (8), and Portability (4). These measures are the result of an analysis of related work [4-7] and the ISO/IEC 25023 [11] and ISO/IEC 9126-3 [14] standards. The relation between the MQuaRE measures and these foundation works can be found in ANNEX B and ANNEX C. Observe that the quality measures presented next can be chosen according to the purpose of the evaluation, the selected quality characteristics, and the possibility of apply the measurements.

### I. COMPLIANCE[3]

#### I.i Conceptual Compliance

Conceptual compliance measures are used to assess the degree to which the conceptual foundation of a metamodel complies with widely accepted and sound theories, regulations, standards, and conventions.

Table 1: Conceptual compliance measures.

| ID | Name | Description | Measurement function |
|---|---|---|---|
| CCc-1 | Conceptual foundation | Which widely-accepted and sound theories, regulations, standards, and conventions is the metamodel | A nominal list of widely-accepted and sound theories, regulations, standards, and conventions to which the metamodel is compliant. |

---
[3] This characteristic was removed from the ISO 25000's quality model, but we included it with specific quality measures for metamodels.

| | | compliant to? | |
|---|---|---|---|
| CCc-2 | Backward Traceability | Which are the metamodel concepts that can be traced back to their conceptual foundations? | A nominal list of each metamodel concept with its respective conceptual foundation. |

## II. CONCEPTUAL SUITABILITY

### II.i Conceptual Completeness

Conceptual completeness measures are used to assess the degree to which the set of metamodel concepts covers all the specified requirements.

Table 2: Conceptual completeness measures.

| ID | Name | Description | Measurement function |
|---|---|---|---|
| CCp-1 | Conceptual coverage | What proportion of the specified concepts has been modeled? | $X = 1 - A / B$<br>A = Number of missing concepts.<br>B = Number of concepts described in the metamodel specification.<br>$0 <= X <= 1$.<br>The closer to 1, the more complete. |
| **NOTE 1**. Concepts can be specified in a metamodel specification, including requirements specification, design specification, user documentation or all of these.<br>**NOTE 2**. A missing concept is detected when the metamodel does not have ability to model a concept that is specified. | | | |

### II.ii Conceptual Correctness

Conceptual correctness measures are used to assess the degree to which the metamodel provides the correct modeling results with the needed degree of precision.

Table 3: Conceptual correctness measures.

| ID | Name | Description | Measurement function |
|---|---|---|---|
| CCr-1 | Conceptual correctness | What proportion of metamodel concepts is modeled correctly? | $X = 1 - A / B$<br>A = Number of incorrectly modeled concepts.<br>B = Number of concepts considered in the evaluation.<br>$0 <= X <= 1$. The closer to 1, the more correct. |

### II.iii Conceptual Appropriateness

Conceptual Appropriateness measures are used to assess the degree to which the metamodel facilitates the accomplishment of modelling tasks, and for determining their adequacy for performing these tasks.

Table 4: Conceptual appropriateness measures.

| ID | Name | Description | Measurement function |
|---|---|---|---|
| CAp-1 | Conceptual appropriateness of usage objective | What proportion of the metamodel concepts provides appropriate outcome to achieve a specific usage objective? | $X = 1 - A / B$<br>A = Number of missing or incorrectly modeled concepts among those that are required for achieving a specific usage objective.<br>B = Number of concepts required for achieving a specific usage objective.<br>$0 <= X <= 1$. The closer to 1, the more appropriateness. |
| **NOTE 1.** This measure will typically be considered for the most important or the most frequently identified usage objectives. Thus, this quality measure is first calculated for each of the defined usage objectives that can be pursued in the metamodel, and then the next quality measure "Conceptual Appropriateness of Metamodel" can be calculated collectively across all usage objectives to provide a metamodel measure. | | | |
| CAp-2 | Conceptual appropriateness of metamodel | What proportion of the metamodel concepts is required by the users to achieve their objectives provides appropriate outcome? | $x = \sum_{i=1 \text{ to } n} A_i / n$<br>$A_i$ = Appropriateness score for usage objective i, |

|   |   |   |   that is, the measured value of CAp-1 for i-th specific usage objective.<br>N = Number of usage objectives.<br>0 <= X <= 1. The closer to 1, the more appropriateness. |
|---|---|---|---|

## III. USABILITY

### III.i Appropriateness Recognizability

Users have to be able to select a metamodel which is suitable for their intended use. The quality measures for appropriateness recognizability are used to assess the degree to which users can recognize whether a metamodel is appropriate for their needs.

Table 5: Appropriateness recognizability measures.

| ID | Name | Description | Measurement function |
|---|---|---|---|
| UAp-1 | Description completeness | What proportion of usage scenarios is described in the metamodel specifications? | X = A / B<br>A = Number of usage scenarios described in the user documents that match usage scenarios described in the metamodel specifications.<br>B = Number of usage scenarios described in the metamodel specifications.<br>0 <= X <= 1. The closer to 1, the more complete. |
| UAp-2 | Demonstration coverage | What proportion of metamodel concepts requiring demonstration have demonstration capability? | X = A / B<br>A = Number of concepts with demonstration features.<br>B = Number of concepts that could benefit from demonstration features.<br>0 <= X <= 1. The closer to 1, the more capable. |
| **NOTE 1.** This measure indicates how much the metamodel specifications demonstrate how the metamodel can be used. This includes "wizards" or "how to". ||||
| UAp-3 | Evident concepts | What proportion of metamodel concepts is evident to the user? | X = A/B<br>A = Number of concepts evident to the user.<br>B = Number of concepts described in the metamodel specification.<br>0 <= X <= 1. The closer to 1, the better. |
| **NOTE 1.** This measure indicates whether users will be able to locate concepts (A) by exploring metamodel specification (B), e.g. by inspecting the metamodel class diagram ||||
| UAp-4 | Concept understandability | What proportion of metamodel concepts is correctly understood without prior training? | X = A / B<br>A = Number of concepts whose purpose is correctly understood without prior training.<br>B = Number of concepts described in the metamodel specification.<br>0 <= X <= 1. The closer to 1, the better. |
| **NOTE 1**. This measure indicates whether users will be able to understand concepts (A) by exploring design specification (e.g. by inspecting the metamodel class diagram). ||||

### III.ii Learnability

Learnability measures are used to assess the degree to which a metamodel can be adopted by specified users to achieve specified goals of learning in a specified context of use.

Table 6: Learnability measures.

| ID | Name | Description | Measurement function |
|---|---|---|---|
| ULe-1 | User guide completeness | What proportion of metamodel concepts is described in the user documentation that enable the use of the metamodel? | X = A / B<br>A = Number of concepts described in the user documentation as required.<br>B = Number of concepts required to be |

| | | documented. |
| | | $0 \leq X \leq 1$. The closer to 1, the more complete. |

**NOTE 1.** Learnability is strongly related to appropriateness recognizability, and appropriateness recognizability measurements are indicators of the learnability potential of the metamodel.

## IV. MAINTAINABILITY

### IV.i Modularity

Modularity measures are used to assess the degree to which a metamodel is composed of discrete concepts such that a change of one concept has minimal impact on other concepts.

Table 7: Modularity measures.

| ID | Name | Description | Measurement function |
|---|---|---|---|
| MMo-1 | Coupling of concepts | How strongly are the concepts independent and how many concepts are free of impacts from changes to other metamodel concepts? | $X = A / B$<br>A = Number of concepts with no impact on others.<br>B = Number of specified concepts which are required to be independent.<br>$0 \leq X \leq 1$. The closer to 1, the less coupling. |
| MMo-2 | Complexity of exercise | How complex is building terminal models by analyzing the structure of the metamodel? | $X = A - B$<br>A = Number of instantiation elements that must be done in order<br>B = Number of instantiation groups that must be completed, but in any order<br>The higher, the more complex, i.e., the metamodel requires more ordered actions when creating the model elements. |
| **NOTE 1**. In the case of hierarchy (specialization/generalization), all created objects inside the hierarchy count as a single instantiation element, when ordered after their parent, whether or not those contained objects are required to be created in a particular order. | | | |

### IV.ii Reusability

Reusability measures are used to assess the degree to which usage scenarios can be used in more than one metamodel.

Table 8: Reusability measures.

| ID | Name | Description | Measurement function |
|---|---|---|---|
| MRe-1 | Reusability per application domain | How reusable is the metamodel to an application domain? | $X = 1 - A / B$<br>A = Number of usage scenarios which were not possible to be reused for an application domain in particular<br>B = Number of usage scenarios described in the metamodel specifications<br>$0 \leq X \leq 1$. The closer to 1, the better. |

### IV.iii Modifiability

Modifiability measures are used to assess the degree to which a metamodel can be effectively and efficiently modified without introducing inconsistencies or degrading existing metamodel quality.

Table 9: Modifiability measures.

| ID | Name | Description | Measurement function |
|---|---|---|---|
| MMd-1 | Conceptual stability | How stable is the metamodel specification during the metamodel's development life cycle? | $X = 1 - A / B$<br>A = Number of concepts changed during the metamodel's development life cycle.<br>B = Number of concepts described in the metamodel specification.<br>$0 \leq X \leq 1$. The closer to 1, the more stable. |
| MMd-2 | Change | Are changes to metamodel | $X = A / B$ |

| | recordability | specifications recorded adequately? | A = Number of changes in concepts having change comments confirmed in review.<br>B = Number of concepts changed from original metamodel specification.<br>0 <= X <= 1. The closer to 1, the more recordable. The change control 0 indicates poor change control. |
|---|---|---|---|
| MMd-3 | Change impact | What is the frequency of adverse impacts after modification? | X = 1 - A / B<br>A = Number of detected adverse impacts after modifications.<br>B = Number of modifications made.<br>0<=X<=1. The closer to 1, the better. |
| MMd-4 | Modification impact localization | How large is the impact of the modification on the metamodel? | X = A / B<br>A = Number of concepts affected by modification, confirmed in review.<br>B = Number of concepts described in the metamodel specification.<br>0 <= X <= 1. The closer to 0, the lesser impact of modification. |
| MMd-5 | Modification correcteness | What proportion of modifications has been implemented correctly? | X = 1 - A / B<br>A = Number of modifications that caused an adverse impact within a defined period after made.<br>B = Number of modifications made.<br>0 <= X <= 1<br>The closer to 1, the better. |

## V. PORTABILITY

### V.i Adaptability

Adaptability measures are used to assess the degree to which a metamodel can effectively and efficiently be adapted for different application domains.

Table 10: Adaptability measures.

| ID | Name | Description | Measurement function |
|---|---|---|---|
| PAd-1 | Adaptability per application domain | How adaptable is the metamodel to an application domain? | X = 1 - A / B<br>A = Number of usage scenarios which were not possible to be modeled for an application domain in particular<br>B = Number of usage scenarios described in the metamodel specifications<br>0 <= X <= 1<br>The closer to 1, the better |

### V.ii Replaceability

Replaceability measures are used to assess the degree to which a metamodel can replace another specified metamodel for the same purpose in the same application domain.

Table 11: Replaceability measures.

| ID | Name | Description | Measurement function |
|---|---|---|---|
| PRe-1 | Usage similarity | What proportion of usage scenarios of the replaced metamodel can be modeled without any additional learning or workaround? | X = A / B<br>A = Number of usage scenarios which can be modeled without any additional learning or workaround<br>B = Number of usage scenarios in the replaced metamodel<br>0 <= X <= 1.<br>The closer to 1, the better. |

| PRe-2 | Metamodel quality equivalence | What proportion of the quality measures is satisfied after replacing previous metamodel by this one? | X = A / B<br>A = Number of quality measures of the new metamodel which are better or equal to the replaced metamodel<br>B = Number of quality measures of the replaced metamodel that are relevant<br>0 <= X <= 1.<br>The closer to 1, the better. |
|---|---|---|---|
| NOTE 1. The relevance of quality measures is specialist's prerogative. | | | |
| PRe-3 | Conceptual inclusiveness | Can the similar concepts easily be used after replacing previous metamodel by this one? | X = A / B<br>A = Number of concepts which produce similar results as before<br>B = Number of concepts which have to be used in the replaced metamodel<br>0 <= X <= 1.<br>The closer to 1, the better. |

## 5. QUALITY EVALUATION PROCESS

The MQuaRE's process model assumes that the evaluation founds on the MQuaRE's requirements, making clear the objectives and criteria of assessment. Besides, the MQuaRE's quality model and measures should also be considered in the evaluation process. Figure 6 depicts a BPMN-based representation for the MQuaRE's evaluation process with activities, user roles, and input and output artifacts. Activities and the respective tasks are detailed next.

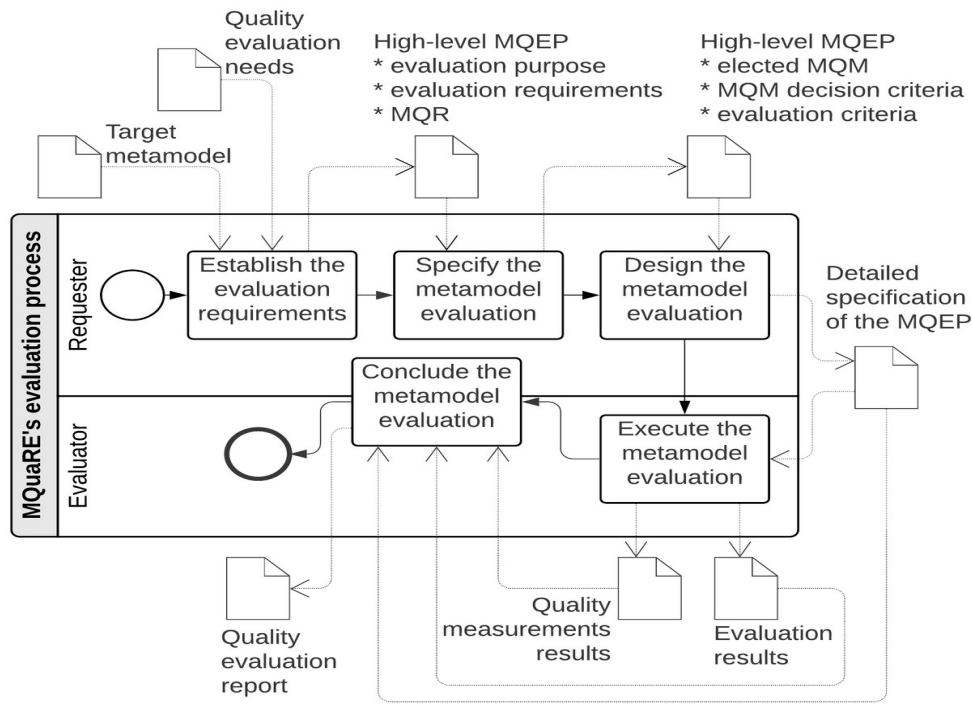

**Figure 6: The MQuaRE's Metamodel Evaluation Process.**

## 5.1. Establish the metamodel evaluation requirements

In this activity, the evaluation requester must identify metamodel quality evaluation requirements, taking into account the evaluation purpose. The inputs for this activity is the metamodel quality evaluation needs; the metamodel to be evaluated, including its specifications, and applicable evaluation tools and methodologies. This activity consists of the following tasks:

### 5.1.1. Establish the objective of evaluating the metamodel

The purpose of the metamodel quality evaluation shall be documented, and will be a basis for the further evaluation activities and tasks. The evaluation purpose depends on the version of the metamodel: final or intermediate. Table 12 shows examples of some evaluation purposes:

### 5.1.2. Define the metamodel quality requirements

The metamodel quality requirements shall be specified using the MQuaRE quality characteristics and sub-characteristics. The evaluation requester can choose the metamodel quality requirements from a preliminary list of 19 (nineteen) metamodel quality requirements that can be reused, reviewed, and refined available and is available in ANNEX D.

### 5.1.3. Identify the artifacts to be used in the evaluation

Every metamodel-related artifact available for the evaluation shall be identified and registered. Examples of metamodel artifacts include: metamodel specifications (requirements and design documents), metamodel implementation, metamodel user documentation, metamodel history documentation, specification of different application domains, metamodel to be replaced by the evaluated metamodel. The availability of some metamodel artifacts will be important to enable the evaluation of some MQRs (see Table 13).

Table 12: Association between metamodel versions and evaluation purposes.

| Metamodel version | Purpose |
|---|---|
| Intermediate version | Assure quality for the metamodel |
| | Decide on the acceptance of an intermediate metamodel version |
| | Access the ongoing feasibility of the ongoing metamodel |
| | Predict or estimate final metamodel quality |
| | Discover improvement points in the metamodel |
| | Collect information on intermediate metamodel version in order to control and manage the process |
| Final version | Decide on the acceptance of the metamodel |
| | Compare a metamodel with others |
| | Select a metamodel from among alternative metamodels |
| | Assess both positive and negative effects of a metamodel |
| | Discover improvement points in the metamodel |

Table 13: Association between metamodel artifacts and metamodel quality requirements.

|  | Metamodel specifications | Metamodel implementation | Metamodel user documentation | Metamodel history documentation | Specification of different application domains | Metamodel to be replaced by the evaluated metamodel |
|---|---|---|---|---|---|---|
| **MQR01** | X |  |  |  |  |  |
| **MQR02** | X | X |  |  |  |  |
| **MQR03** | X | X |  |  |  |  |
| **MQR04** | X | X | X |  |  |  |
| **MQR05** | X |  | X |  |  |  |
| **MQR06** | X |  | X |  |  |  |
| **MQR07** | X | X |  |  |  |  |
| **MQR08** | X | X |  |  |  |  |
| **MQR09** | X |  | X |  |  |  |
| **MQR10** |  | X |  |  |  |  |
| **MQR11** |  | X |  |  |  |  |
| **MQR12** | X |  | X |  | X |  |
| **MQR13** | X |  |  | X |  |  |
| **MQR14** | X |  |  | X |  |  |
| **MQR15** | X |  |  | X |  |  |
| **MQR16** |  |  | X |  | X |  |
| **MQR17** |  |  | X |  |  | X |
| **MQR18** | X | X |  |  |  | X |
| **MQR19** | X | X |  |  |  |  |

As illustrated in Figure 6, the main output artifact of this activity "Establish the metamodel evaluation requirements" is a high-level Metamodel Quality Evaluation Plan (MQEP – a template for the MQEP is available in ANNEX I). A first version of the MQEP should contain the purposes of the metamodel evaluation, the specification of metamodel quality requirements, and the artifacts available to be used in the evaluation execution.

## 5.2. Specify the metamodel evaluation

This activity is executed by the evaluation requester. It consumes a high-level MQEP as input and consists of the following tasks:

### 5.2.1. Select the metamodel quality measures

The requester shall select quality measures to cover all MQRs chosen in Section 5.1.2. The ANNEX E presents the list of 23 (twenty three) metamodel quality measures organized by requirements that could be chosen according to the requirements selected for the evaluation.

### 5.2.2. Define decision criteria for metamodel quality measures

Decision criteria are numerical thresholds or targets used to determine the need for action or further investigation or to describe the level of confidence in a given result. The requester must define target value and acceptance tolerance for each selected measure.

Figure 7 presents an example with the measure "conceptual coverage" where the target value was set to 1, i.e., full conceptual coverage of the metamodel's implementation in relation to its specifications. In addition, an acceptable value of 0.75 is also assigned, i.e., quality requirement 2 can be met if at least 75% of the specified concepts are implemented. Although the target and tolerance values are the prerogative of the evaluation requester, ANNEX F offers a template to facilitate the definition of these values for each selected measure.

Table 14: The example of decision criteria for conceptual coverage measure.

| Characteristic | Sub-characteristic | MQR | Measures | Measure Description | Interpretation of the measurement value | Target value | Acceptable tolerance value |
|---|---|---|---|---|---|---|---|
| Conceptual suitability | Conceptual completeness | MQR02 | CCp-1 - Conceptual coverage | What proportion of the specified concepts has been modeled? | The closer to 1, the more complete. | 1 | 0,75 |

### 5.2.3. Establish decision criteria for evaluating the metamodel

The requester should prepare a procedure for further summarization, with separate criteria for different quality characteristics, each of which may be in terms of individual quality sub-characteristics and measures. The formulas must be defined according to the notes received in each individual quality measure.

Figure 8 presents an example, in which the sub-characteristic Conceptual Apropriateness has 2 measures (CAp-1 and CAp-2) and the general score for this sub-characteristic is given by the arithmetic mean of CAp-1 and CAp-2. Note that the general mark of the characteristic in question is given by the arithmetic average of the marks of each sub-characteristic of quality. It is worth mentioning that although these formulas are the requester's choice, ANNEX G provides a form to define these formula to calc the quality value for characteristics and sub-characteristics.

Table 15: The example of decision criteria for evaluating the metamode.l

| Characteristic | Sub-characteristic | Quality Requirements | Measures | Sub-characteristic formula | Characteristic formula |
|---|---|---|---|---|---|
| Conceptual suitability | Conceptual completeness | MQR02 - The metamodel must cover the concepts found in its specifications. | CCp-1 - Conceptual coverage | CCp1 | (CCp1 + CCr1 + CAp) / 3 |
| | Conceptual correctness | MQR03 - The metamodel must represent the concepts found in its specifications correctly. | CCr-1 - Conceptual correctness | CCr1 | |
| | Conceptual appropriateness | MQR04 - The metamodel must represent the concepts required for achieving specific usage objectives. | CAp-1 - Conceptual appropriateness of usage objective | CAp = (CAp1 + CAp2) / 2 | |
| | | | CAp-2 - Conceptual appropriateness of metamodel | | |

The primary output artifact of this activity is a revised high-level MQEP, containing the chosen quality measures as well as decision criteria for metamodel quality measures and assessment.

## 5.3. Design the metamodel evaluation

This is the last activity executed by the evaluation requester before starting the evaluation execution. The revised high-level MQEP previously presented is input data for this activity, and a evaluation plan shall be defined.

### 5.3.1. Plan metamodel evaluation activities

In this task, those metamodel quality evaluation activities identified shall be scheduled, taking into account the availability of resources such as personnel, evaluation tools, and examples of metamodel application domains.

The evaluation plan should be documented with the following elements:
- The purpose of the metamodel quality evaluation;
- Quality evaluation requirements, including
    - the metamodel artifacts to be used in the evaluation;
    - evaluation resources (like as personnel, tools, budget, and deadlines);
- The metamodel quality requirements;
- The metamodel quality measures;
- Decision criteria for metamodel evaluation and metamodel quality measures;
- An evaluation schedule.

As the evaluation activities evolve, the evaluation plan shall be revised until a thorough level plan. The outcome of this activity is a detailed specification of the MQEP (a template for the MQEP is available in ANNEX I).

### 5.4. Execute the metamodel evaluation

This activity starting the metamodel quality evaluation in which the evaluator use the thorough specification of the MQEP as the input artifact for the following tasks:

#### 5.4.1. Compute metamodel quality measurements

The selected metamodel quality measures and described in the MQEP shall be applied to the metamodel. As a result, values on the measurement scales are computed and assigned to quality measures.

#### 5.4.2. Apply decision criteria for metamodel quality measures

The decision criteria for measures defined in the MQEP (task described in Section 5.2.2) shall be applied to the measured values.

#### 5.4.3. Apply decision criteria for metamodel evaluation

The set of decision criteria for assessment defined in the MQEP (task described in Section 5.2.3) shall be summarized into quality characteristics and sub-characteristics. A statement of the extent to which the metamodel meets quality requirements describes the assessment results, which should:

1. establish an appropriate degree of confidence that the metamodel can meet the evaluation requirements;
2. identify specific deficiencies concerning the evaluation requirements and additional evaluations needed to determine the scope of those deficiencies;
3. identify particular limitations or conditions placed on the use of the metamodel;
4. identify weaknesses or omissions in the evaluation and any additional evaluation that is needed.

The outcomes of this activity are the: (a) measured value, (b) final measure value, (c) sub-characteristic value, and (d) characteristic value. The final measurement value is the measured value compared to the target value and acceptance value defined in the activity "5.2.3. Establish decision criteria for evaluating the metamodel". The evaluator must document all of these values. For this, ANNEX H offers a measurement table that can ve used to facilitate this task.

Table 16: The example of measurement table.

| Characteristic | Sub-characteristic | Quality Requirement | Measure | Measured value | Final measurement value | Sub-characteristic value | Characteristic value |
|---|---|---|---|---|---|---|---|
| Conceptual suitability | Conceptual completeness | MQR02 - The metamodel must cover the concepts found in its specifications. | Ccp-1 Conceptual coverage | A = 0<br>B = 20<br>x = 1 | 1.0 | 1.0 | |

### 5.5. Conclude the metamodel evaluation

This activity requires as input the detailed specification of MQEP, metamodel quality measurements results, and quality evaluation results. It consists of the following tasks:

#### 5.5.1. Review the metamodel evaluation results

Both the metamodel evaluator and the evaluation requester shall carry out a joint review of the evaluation results. All documentation generated must be reassessed, and adaptations can be made when justified and documented.

#### 5.5.2. Create the metamodel evaluation report

Depending on how the evaluation report is to be used, it should include the following items, among others: the MQEP, computed measurements results, performed analyses, intermediate results or interpretation decisions, the evaluators' profiles, the final result of the metamodel quality evaluation, and any necessary information to be able to repeat or reproduce the assessment.

The final outcome is a metamodel quality evaluation report (a template for the report is available in Annex J).

## 6. FINAL REMARKS

The MQuaRE framework was proposed as a response to the lack of a more thorough solution for metamodel quality evaluation. The MQuaRE requirements, model, and measures can be adapted to the specific contexts of a metamodel. New characteristics and sub-characteristics can be included in the MQuaRE's quality model as well as corresponding measures and requirements. The MQuaRE framework should serve as a guide for metamodel quality evaluation, but it should not hamper execution.

## 7. REFERENCES

[1] Markus Herrmannsdorfer and Guido Wachsmuth. 2014. Evolving Software Systems. Springer, Berlin, Heidelberg, Chapter Coupled Evolution of Software Metamodels and Models, 33–63.

[2] OMG. 2002. Meta Object Facility (MOF) Specification, Version 1.4. Object Management Group, Inc. (2002).

[3] Éric Vépa, Jean Bézivin, Hugo Bruneliere, and Frédéric Jouault. 2006. Measuring model repositories. In Model Size Metrics Workshop - a MODELS 2006 Satellite Event. Springer, Genoa, Italy, 1–5.

[4] Haohai Ma, Weizhong Shao, Lu Zhang, Zhiyi Ma, and Yanbing Jiang. 2004. Applying OO Metrics to Assess UML Meta-models. In UML 2004 — The Unified Modeling Language. Modeling Languages and Applications, Thomas Baar, Alfred Strohmeier, Ana Moreira, and Stephen J. Mellor (Eds.). Springer, Berlin, 12–26.

[5] Vjeran Strahonja. 2007. The Evaluation Criteria of Workflow Metamodels. In 29th International Conference on Information Technology Interfaces. IEEE, New York, NY, USA, 553–558.

[6] Zhiyi Ma, Xiao He, and Chao Liu. 2013. Assessing the quality of metamodels. Frontiers of Computer Science 7, 4, Article 558 (2013), 12 pages.

[7] Juri Rocco, Davide Di Ruscio, Ludovico Iovino, and Alfonso Pierantonio. 2014. Mining metrics for understanding metamodel characteristics. In Proceedings of the 6th International Workshop on Modeling in Software Engineering (MiSE 2014). ACM, New York, NY, USA, 55–60.

[8] James Williams, Athanasios Zolotas, Nicholas Matragkas, Louis Rose, Dimitios Kolovos, Richard Paige, and Fiona Polack. 2013. What do metamodels really look like?. In Proceedings of the 3rd International Workshop on Experiences and Empirical Studies in Software Modeling. CEUR-WS, 55–60.

[9] ISO/IEC. 2014. ISO/IEC 25000:2014 Systems and software engineering — Systems and software Quality Requirements and Evaluation (SQuaRE) — Guide to SQuaRE. ISO/IEC 25000:2014 2 (2014), 1–27

[10] ISO/IEC. 2011. ISO/IEC 25040:2011. Systems and software engineering — Systems and software Quality Requirements and Evaluation (SQuaRE) — Evaluation process.

[11] ISO/IEC. 2016. ISO/IEC 25023:2016 Systems and software engineering — Systems and software Quality Requirements and Evaluation (SQuaRE) — Measurement of system and software product quality. ISO/IEC 25023:2016 1 (2016), 1–45.

[12] ISO/IEC. 2011. ISO/IEC 25010:2011 Systems and software engineering — Systems and software Quality Requirements and Evaluation (SQuaRE) — System and software quality models.

[13] ISO/IEC. 2001. ISO/IEC 9126-1:2001 Software engineering — Product quality — Part 1: Quality model. ISO/IEC 9126-1:2001 1 (2001), 1–25.

[14] ISO/IEC. 2003. ISO/IEC TR 9126-3:2003 Software engineering — Product quality — Part 3: Internal metrics. ISO/IEC 9126-3:2003 2 (2003), 1–62.

[15] Strahonja, V. "The Evaluation Criteria of Workflow Metamodels," 2007 29th International Conference on Information Technology Interfaces, Cavtat, 2007, pp. 553-558, doi: 10.1109/ITI.2007.4283831

[16] Ma et al. "Applying OO Metrics to Assess UML Meta-models. UML - The Unified Modeling Language. Modeling Languages and Applications". 2004. Springer Berlin Heidelberg, Berlin, Heidelberg, 12–26. DOI:https://doi.org/10.1007/978-3-540-30187-5_2

[17] Ma, Z., He, X. & Liu, C. "Assessing the quality of metamodels". 2013. Front. Comput. Sci. 7, 558–570. https://doi.org/10.1007/s11704-013-1151-5

[18] Sprinkle, J. "Analysis of a metamodel to estimate complexity of using a domain-specific language". 2010. In Proceedings of the 10th Workshop on Domain-Specific Modeling (DSM '10). Association for Computing Machinery, New York, NY, USA, Article 13, 1–6. DOI:https://doi.org/10.1145/2060329.2060359

# ANNEX A – COMPARISON WITH THE PREVIOUS MODELS

| MQuaRE Model | ISO/IEC 25010 [12] | ISO/IEC 9126-1 [13] | Related work |
|---|---|---|---|
| **1. Compliance** | | | |
| 1.1. Conceptual compliance | | Functionality compliance | [15] |
| **2. Conceptual suitability** | **Functional suitability** | **Functionality** | [16] [17] |
| 2.1. Conceptual completeness | Functional completeness | | [15] |
| 2.2. Conceptual correctness | Functional correctness | Accuracy | [15] |
| 2.3. Conceptual appropriateness | Functional appropriateness | Suitability | |
| **3. Usability** | **Usability** | | |
| 3.1. Appropriateness recognizability | Appropriateness recognizability | Understandability | [15] [16] [17] |
| 3.2. Learnability | Learnability | Learnability | |
| **4. Maintainability** | **Maintainability** | **Maintainability** | |
| 4.1. Modularity | Modularity | | [15] [18] |
| 4.2. Reusability | Reusability | | [16] [17] |
| 4.3. Modifiability | Modifiability | Stability | [15] |
| **5. Portability** | **Portability** | **Portability** | |
| 5.1. Adaptability | Adaptability | Adaptability | [15] |
| 5.2. Replaceability | Replaceability | Replaceability | [15] |

# ANNEX B – ORIGINS OF MQUARE MEASURES

The following list presents 23 measures: three new ones, twelve adapted from the ISO/IEC 25023 [11], seven adapted from the ISO/IEC 9126-3 [14], and one from a research paper.

| MQuaRE measures | ISO/IEC 25023:2016  [11] |
|---|---|
| **Compliance measures (NEW)** | |
|   Conceptual compliance measures **(NEW)** | |
|     CCc-1 Conceptual foundation **[14]** <br>     CCc-2 Backward Traceability **(NEW)** | |
|   Conceptual suitability measures |   Functional suitability measures |
|     Conceptual completeness measures |     Functional completeness measures |
|       Ccp-1 Conceptual coverage |       Fcp-1-G Functional coverage |
|     Conceptual correctness measures |     Functional correctness measures |
|       CCr-1 Conceptual correctness |       FCr-1-G Functional correctness |
|     Conceptual appropriateness measures |     Functional appropriateness measures |
|       CAp-1 Conceptual appropriat. of usage objective <br>       CAp-2 Conceptual appropriat. of metamodel |       FAp-1-G Functional  appropriat.of usage objective <br>       FAp-2-G Functional appropriat. of system |
| Usability measures | Usability measures |
|   Appropriateness recognizability measures |   Appropriateness recognizability measures |
|     UAp-1 Description completeness <br>     UAp-2 Demonstration coverage <br>     UAp-3 Evident concepts **[14]** <br>     UAP-4 Concept understandability **[14]** |     UAp-1-G Description completeness <br>     UAp-2-S Demonstration coverage |
|   Learnability measures |   Learnability measures |
|     ULe-1 User guide completeness |     ULe-1-G User guide completeness |
| **Maintainability measures** | **Maintainability measures** |
|   Modularity measures |   Modularity measures |
|     MMo-1 Coupling of concepts <br>     MMo-2 Complexity of exercise **[18]** |     MMo-1-G Coupling of concepts |
|   Reusability measures |   Reusability measures |
|     MRe-1 Reusability per application domain **(NEW)** | |
|   Modifiability measures |   Modifiability measures |
|     MMd-1 Conceptual stability **[14]** <br>     MMd-2 Change recordability **[14]** <br>     MMd-3 Change impact **[14]** <br>     MMd-4 Modific. impact localization **[14]** <br>     MMd-5 Modification correctness |     MMd-3-S Modification correctness |
| **Portability measures** | **Portability measures** |
|   Adaptability measures |   Adaptability measures |
|     PAd-1 Adaptability per application domain **(NEW)** | |
|   Replaceability measures <br>     PRe-1 Usage similarity <br>     PRe-2 Metamodel quality equivalence <br>     PRe-3 Conceptual inclusiveness |   Replaceability measures <br>     PRe-1-G Usage similarity <br>     PRe-2-S Product quality equivalence <br>     PRe-3-S Functional  inclusiveness |

# ANNEX C - MEASURES OF ISO/IEC 25023:2016 INCLUDED IN MQUARE

The following list presents the measures of ISO/IEC 25023:2016 [11] included and not included in MQuaRE due to the understanding of not being part of the scope of the quality of a metamodel. Also shown are the new measures that were defined in MQuaRE that complements the list of quality measures.

| ISO/IEC 25023:2016 [11] | MQuaRE measures |
|---|---|
| | Compliance measures **(NEW)** |
| |     Conceptual compliance measures **(NEW)** |
| |         CCc-1 Conceptual foundation **[14]** |
| |         CCc-2 Backward traceability **(NEW)** |
| Functional suitability measures | Conceptual suitability measures |
|     Functional completeness measures |     Conceptual completeness measures |
|         Fcp-1-G Functional coverage |         CCp-1 Conceptual coverage |
|     Functional correctness measures |     Conceptual correctness measures |
|         FCr-1-G Functional correctness |         CCr-1 Conceptual correctness |
|     Functional appropriateness measures |     Conceptual appropriateness measures |
|         FAp-1-G Functional appropriat.of usage objective |         CAp-1 Conceptual appropriat. of usage objective |
|         FAp-2-G Functional appropriat. of system |         CAp-2 Conceptual appropriat. of metamodel |
| Performance efficiency measures | NAM[4] |
|     Time behaviour measures | |
|         PTb-1-G Mean response time | |
|         PTb-2-G Response time adequacy | |
|         PTb-3-G Mean turnaround time | |
|         PTb-4-G Turnaround timea dequacy | |
|         PTb-5-G Mean throughput | |
|     Resource utilization measures | |
|         PRu-1-G Mean processor utilization | |
|         PRu-2-G Mean memory utilization | |
|         PRu-3-G Mean I/O devices utilization | |
|         PRu-4-S Bandwidth utilization | |
|     Capacity measures | |
|         PCa-1-G Transaction processing capacity | |
|         PCa-2-G User access capacity | |
|         PCa-3-S User access increase adequacy | |
| Compatibility measures | NAM |
|     Co-existence measures | |
|         Cco-1-G Co-existence with other products | |
|     Interoperability measures | |
|         Cin-1-G Data formats exchangeability | |
|         Cin-2-G Data exchange protocol sufficiency | |
|         Cin-3-S External interface adequacy | |
| Usability measures | Usability measures |
|     Appropriateness recognizability measures |     Appropriateness recognizability measures |
|         UAp-1-G Description completeness |         UAp-1 Description completeness |
|         UAp-2-S Demonstration coverage |         UAp-2 Demonstration coverage |
|         UAp-3-S Entry point self-descriptiness |         UAp-3 Evident concepts **[14]** |
| |         UAP-4 Concept understandability **[14]** |
|     Learnability measures |     Learnability measures |
|         ULe-1-G User guide completeness |         ULe-1 User guide completeness |
|         ULe-2-S Entry fields defaults | |
|         ULe-3-S Error message understandability | |
|         ULe-4-S Self-explanatory user interface | |

---

[4] These characteristics and sub-characteristics do not apply to metamodels (NAM), in our humble opinion.

| ISO/IEC 25023:2016 [11] | MQuaRE measures |
|---|---|
| Operability | NAM |
|     UOp-1-G Operational consistency | |
|     UOp-2-G Message clarity | |
|     UOp-3-S Functional customizability | |
|     UOp-4-S User interface customizability | |
|     UOp-5-S Monitoring capability | |
|     UOp-6-S Undo capability | |
|     UOp-7-S Understandable categorization of inform. | |
|     UOp-8-S Appearance consistency | |
|     UOp-9-S Input device support | |
| User error protection measures | NAM |
|     UEp-1-G Avoidance of user operation error | |
|     UEp-2-S User entry error correction | |
|     UEp-3-S User error recoverability | |
| User interface aesthetics measures | NAM |
|     UIn-1-S Appearance aesthetics of user interfaces | |
| Accessibility measures | NAM |
|     UAc-1-G Accessibility for users with disabilities | |
|     UAc-2-S Supported languages adequacy | |
| Reliability measures | NAM |
|   Maturity measures | |
|     RMa-1-G Fault correction | |
|     RMa-2-G Mean Time between failure | |
|     RMa-3-G Failure rate | |
|     RMa-4-S Test coverage | |
|   Availability measures | |
|     RAv-1-G System availability | |
|     RAv-2-G Mean down time | |
|   Fault tolerance measures | |
|     RFt-1-G Failure avoidance | |
|     RFt-2-S Redundancy of components | |
|     RFt-3-S Mean fault notification time | |
|   Recoverability measures | |
|     RRe-1-G Mean recovery time | |
|     RRe-2-S Backup data completeness | |
| Security measures | NAM |
|   Confidentiality measures | |
|     SCo-1-G Access controllability | |
|     SCo-2-G Data encryption correcteness | |
|     SCo-3-S Strength of cryptographic algorithms | |
|   Integrity measures | |
|     Sin-1-G Data integrity | |
|     Sin-2-G Internal data corruption prevention | |
|     Sin-3-S Buffer overflow prevention | |
|   Non-repudiation measures | |
|     SNo-1-G Digital signature usage | |
|   Accountability measures | |
|     SAc-1-G User audit trail completeness | |
|     SAc-2-S System log retention | |
|   Authenticity measures | |
|     SAu-1-G Authentication mechanism sufficiency | |
|     SAu-2-S Authentication rules conformity | |
| Maintanability measures | Maintanability measures |
|   Modularity measures |   Modularity measures |
|     MMo-1-G Coupling of concepts |     MMo-1-G Coupling of concepts |
|     MMo-2-S Cyclomatic complexity adequacy |     MMo-2 Complexity of exercise **[Sprinkle,2010]** |

| ISO/IEC 25023:2016 [11] | MQuaRE measures |
|---|---|
| Reusability measures<br>    MRe-1-G Reusability of assets<br>    MRe-2-S Coding rules conformity | Reusability measures<br>    MRe-1 Reusability per application domain **(NEW)** |
| Analisability measures<br>    MAn-1-G System log completeness<br>    MAn-2-S Diagnosis function effectiveness<br>    MAn-3-S Diagnosis function sufficiency | NAM |
| Modifiability measures<br>    MMd-1-G Modification efficiency<br>    MMd-2-G Modification correctness<br>    MMd-3-S Modification capability | Modifiability measures<br>    MMd-1 Conceptual stability **[14]**<br>    MMd-2 Change recordability **[14]**<br>    MMd-3 Change impact **[14]**<br>    MMd-4 Modific. impact localization **[14]**<br>    MMd-5 Modification correctness |
| Testability measures<br>    MTe-1-G Test function completeness<br>    MTe-2-S Autonomous testability<br>    MTe-3-S Test restartability | NAM |
| Portability measures | Portability measures |
|   Adaptability measures<br>    PAd-1-G Hardware environmental adaptability<br>    PAd-2-G System softw. environmental adaptability<br>    PAd-3-S Operational environmental adaptability |   Adaptability measures<br>    PAd-1 Adaptability per application domain **(NEW)** |
|   Installability measures<br>    PIn-1-G Installation time efficiency<br>    PIn-2-G Ease of installation | NAM |
|   Replaceability measures<br>    PRe-1-G Usage similarity<br>    PRe-2-S Product quality equivalence<br>    PRe-3-S Functional inclusiveness<br>    PRe-4-S Data reusability/import capability |   Replaceability measures<br>    PRe-1 Usage similarity<br>    PRe-2 Metamodel quality equivalence<br>    PRe-3 Conceptual inclusiveness |

# ANNEX D – METAMODEL QUALITY REQUIREMENTS

| Characteristic | Sub-characteristic | Metamodel Quality Requirements |
|---|---|---|
| Compliance | Conceptual compliance | MQR01 - The metamodel conceptual foundation must comply with widely-accepted and sound theories, regulations, standards, and conventions. |
| Conceptual suitability | Conceptual completeness | MQR02 - The metamodel must cover the concepts found in its specifications. |
| | Conceptual correctness | MQR03 - The metamodel must represent the concepts found in its specifications correctly. |
| | Conceptual appropriateness | MQR04 - The metamodel must represent the concepts required for achieving specific usage objectives. |
| Usability | Appropriateness recognizability | MQR05 - The users must be able to recognize whether a metamodel is appropriate for their needs accordingly the usage scenarios described in the user documents. |
| | | MQR06 - The users must be able to recognize whether a metamodel is appropriate for their needs accordingly the demonstration features of metamodel concepts. |
| | | MQR07 - The users must be able to recognize whether a metamodel is appropriate for their needs accordingly the evident concepts to the user in the metamodel specifications. |
| | | MQR08 - The users must be able to recognize whether a metamodel contain concepts whose purpose is correctly understood without prior training. |
| | Learnability | MQR09 - The users must be able to recognize whether a metamodel is appropriate for their needs accordingly the metamodel user documentation. |
| Maintainability | Modularity | MQR10 - The metamodel must be composed of discrete concepts such that a change of one concept has minimal impact on other concepts. |
| | | MQR11 - The metamodel must be composed of discrete concepts such that a creation of model elements does not enforce ordered modelling actions. |
| | Reusability | MQR12 - The metamodel must be able to be reused to modelling usage scenarios for different application domains. |
| | Modifiability | MQR13 - The users must be able to recognize metamodel modifications acoordingly the changes documented in the metamodel specification during metamodel development life cycle. |
| | | MQR14 - The users must be able to recognize metamodel modifications acoordingly the change comments confirmed in review. |
| | | MQR15 - The metamodel must be reused modified without introducing inconsistencies or degrading metamodel quality. |

| Characteristic | Sub-characteristic | Metamodel Quality Requirements |
|---|---|---|
| Portability | Adaptability | MQR16 - The metamodel must be able to be adapted to modelling usage scenarios for different application domains. |
| | Replaceability | MQR17 - The metamodel must be able to replace another specified metamodel for the same purpose in the same application domain, without introducing any additional learning or workaround. |
| | | MQR18 - The metamodel must be able to replace another specified metamodel for the same purpose in the same application domain, without degrading metamodel quality degree. |
| | | MQR19 - The metamodel must be able to replace another specified metamodel for the same purpose in the same application domain by using similar concepts of previous metamodel. |

# ANNEX E - MQUARE QUALITY MEASURES

Metamodel quality measures organized by characteristics, sub-characteristic, and metamodel quality requirements.

| Characteristic | Sub-characteristic | MQR | Measures | Measure Description | Measurement function |
|---|---|---|---|---|---|
| Compliance | Conceptual compliance | MQR01 | CCc-1 - Conceptual foundation | Which widely-accepted and sound theories, regulations, standards, and conventions is the metamodel compliant to? | A nominal list of widely-accepted and sound theories, regulations, standards, and conventions to which the metamodel is compliant. |
| | | | CCc-2 - Backward Traceability | Which are the metamodel concepts that can be traced back to their conceptual foundations? | A nominal list of each metamodel concept with its respective conceptual foundation. |
| Conceptual suitability | Conceptual completeness | MQR02 | CCp-1 - Conceptual coverage | What proportion of the specified concepts has been modeled? | $X = 1 - A / B$<br>A = Number of missing concepts.<br>B = Number of concepts described in the metamodel specification.<br>$0 <= X <= 1$.<br>The closer to 1, the more complete. |
| | Conceptual correcteness | MQR03 | CCr-1 - Conceptual correcteness | What proportion of metamodel concepts are modeled correctly? | $X = 1 - A / B$<br>A = Number of incorrectly modeled concepts.<br>B = Number of concepts considered in the evaluation.<br>$0 <= X <= 1$.<br>The closer to 1, the more correct. |
| | Conceptual appropriateness | MQR04 | CAp-1 - Conceptual appropriateness of usage objective | What proportion of the metamodel concepts provides appropriate outcome to achieve a specific usage objective? | $X = 1 - A / B$<br>A = Number of missing or incorrectly modeled concepts among those that are required for achieving a specific usage objective.<br>B = Number of concepts required for achieving a specific usage objective.<br>$0 <= X <= 1$.<br>The closer to 1, the more appropriateness. |
| | | | CAp-2 - Conceptual appropriateness of metamodel | What proportion of the metamodel concepts required by the users to achieve their objectives provides appropriate outcome? | $$x = \sum_{i=1 \text{ to } n} A_i / n$$<br>Ai = Appropriateness score for usage objective i, that is, the measured value of CAp-1 for i-th specific usage objective.<br>N = Number of usage objectives.<br>The closer to 1, the more appropriateness. |

| Characteristic | Sub-characteristic | MQR | Measures | Measure Description | Measurement function |
|---|---|---|---|---|---|
| Usability | Appropriateness recognizability | MQR05 | UAp-1 - Description completeness | What proportion of usage scenarios is described in the metamodel specifications? | X = A / B<br>A = Number of usage scenarios described in the user documents that match usage scenarios described in the metamodel specifications.<br>B = Number of usage scenarios described in the metamodel specifications.<br>0 <= X <= 1.<br>The closer to 1, the more complete. |
| | | MQR06 | UAp-2 - Demonstration coverage | What proportion of metamodel concepts requiring demonstration have demonstration capability? | X = A / B<br>A = Number of concepts with demonstration features.<br>B = Number of concepts that could benefit from demonstration features.<br>0 <= X <= 1.<br>The closer to 1, the more capable. |
| | | MQR07 | UAp-3 - Evident concepts | What proportion of metamodel concepts are evident to the user? | X = A/B<br>A = Number of concepts evident to the user.<br>B = Number of concepts described in the metamodel specification.<br>0 <= X <= 1.<br>The closer to 1, the better. |
| | | MQR08 | UAp-4 - Concept understandability | What proportion of metamodel concepts are correctly understood without prior training? | X = A / B<br>A = Number of concepts whose purpose is correctly understood without prior training.<br>B = Number of concepts described in the metamodel specification.<br>0 <= X <= 1.<br>The closer to 1, the better. |
| | Learnability | MQR09 | ULe-1 - User guide completeness | What proportion of metamodel concepts is described in the user documentation that enable the use of the metamodel? | X = A / B<br>A = Number of concepts described in the user documentation as required.<br>B = Number of concepts required to be documented.<br>0 <= X <= 1.<br>The closer to 1, the more complete. |

| Characteristic | Sub-characteristic | MQR | Measures | Measure Description | Measurement function |
|---|---|---|---|---|---|
| Maintainability | Modularity | MQR10 | MMo-1 - Coupling of concepts | How strongly are the concepts independent and how many concepts are free of impacts from changes to other metamodel concepts? | X = A / B<br>A = Number of concepts with no impact on others<br>B = Number of specified concepts which are required to be independent.<br>0 <= X <= 1.<br>The closer to 1, the less coupling. |
| Maintainability | Modularity | MQR11 | MMo-2 - Complexity of exercise | How complex is building terminal models by analyzing the structure of the metamodel? | X = A − B<br>A = Number of instantiation elements that must be done in order<br>B = Number of instantiation groups that must be completed, but in any order<br>The higher, the more complex, i.e., the metamodel requires more ordered actions when creating the model elements. |
| Maintainability | Reusability | MQR12 | MRe-1 - Reusability per application domain | How reusable is the metamodel to an application domain? | X = 1 - A / B<br>A = Number of usage scenarios which were not possible to be reused for an application domain in particular<br>B = Number of usage scenarios described in the metamodel specifications<br>0 <= X <= 1.<br>The closer to 1, the better. |
| Maintainability | Modifiability | MQR13 | MMd-1 - Conceptual stability | How stable is the metamodel specification during the metamodel's development life cycle? | X = 1 - A / B<br>A = Number of concepts changed during the metamodel's development life cycle.<br>B = Number of concepts described in the metamodel specification.<br>0 <= X <= 1<br>The closer to 1, the more stable. |
| Maintainability | Modifiability | MQR14 | MMd-2 - Change recordability | Are changes to metamodel specifications recorded adequately? | X = A / B<br>A = Number of changes in concepts having change comments confirmed in review.<br>B = Number of concepts changed from original metamodel specification.<br>0 <= X <= 1.<br>The closer to 1, the more recordable.<br>The change control 0 indicates poor change control. |
| | | MQR15 | MMd-3 - Change impact | What is the frequency of adverse impacts after modification? | X = 1 - A / B<br>A = Number of detected adverse impacts after modifications.<br>B = Number of modifications made.<br>0 <= X <= 1<br>The closer to 1, the better. |

| Characteristic | Sub-characteristic | MQR | Measures | Measure Description | Measurement function |
|---|---|---|---|---|---|
| | | | MMd-4 - Modification impact localization | How large is the impact of the modification on the metamodel? | X = A / B<br>A = Number of concepts affected by modification, confirmed in review.<br>B = Number of concepts described in the metamodel specification.<br>0 <= X <= 1.<br>The closer to 0, the lesser impact of modification. |
| | | | MMd-5 - Modification correctness | What proportion of modifications has been implemented correctly? | X = 1 - A / B<br>A = Number of modifications that caused an adverse impact within a defined period after made.<br>B = Number of modifications made.<br>0 <= X <= 1<br>The closer to 1, the better. |
| Portability | Adaptability | MQR16 | PAd-1 - Adaptability per application domain | How adaptable is the metamodel to an application domain? | X = 1 - A / B<br>A = Number of usage scenarios which were not possible to be modeled for an application domain in particular<br>B = Number of usage scenarios described in the metamodel specifications<br>0 <= X <= 1<br>The closer to 1, the better |
| | Replaceability | MQR17 | PRe-1 - Usage similarity | What proportion of usage scenarios of the replaced metamodel can be modeled without any additional learning or workaround? | X = A / B<br>A = Number of usage scenarios which can be modeled without any additional learning or workaround<br>B = Number of usage scenarios in the replaced metamodel<br>0 <= X <= 1.<br>The closer to 1, the better. |
| | | MQR18 | PRe-2 - Metamodel quality equivalence | What proportion of the quality measures is satisfied after replacing previous metamodel by this one? | X = A / B<br>A = Number of quality measures of the new metamodel which are better or equal to the replaced metamodel<br>B = Number of quality measures of the replaced metamodel that are relevant<br>0 <= X <= 1.<br>The closer to 1, the better. |
| | | MQR19 | PRe-3 - Conceptual inclusiveness | Can the similar concepts easily be used after replacing previous metamodel by this one? | X = A / B<br>A = Number of concepts which produce similar results as before<br>B = Number of concepts which have to be used in the replaced metamodel<br>0 <= X <= 1.<br>The closer to 1, the better. |

# ANNEX F – METAMODEL QUALITY REQUIREMENTS, MEASURES, ARTIFACTS AND TARGET VALUES

| Characteristic | Sub-characteristic | MQR | Measures | Measure Description | Interpretation of the measurement value | Target value | Acceptable tolerance value |
|---|---|---|---|---|---|---|---|
| Compliance | Conceptual compliance | MQR01 | CCc-1 - Conceptual foundation | Which widely-accepted and sound theories, regulations, standards, and conventions is the metamodel compliant to? | | | |
| | | | CCc-2 - Backward Traceability | Which are the metamodel concepts that can be traced back to their conceptual foundations? | | | |
| Conceptual suitability | Conceptual completeness | MQR02 | CCp-1 - Conceptual coverage | What proportion of the specified concepts has been modeled? | The closer to 1, the more complete. | | |
| | Conceptual correctness | MQR03 | CCr-1 - Conceptual correctness | What proportion of metamodel concepts are modeled correctly? | The closer to 1, the more correct. | | |
| | Conceptual appropriateness | MQR04 | CAp-1 - Conceptual appropriateness of usage objective | What proportion of the metamodel concepts provides appropriate outcome to achieve a specific usage objective? | The closer to 1, the more appropriateness. | | |
| | | | CAp-2 - Conceptual appropriateness of metamodel | What proportion of the metamodel concepts required by the users to achieve their objectives provides appropriate outcome? | The closer to 1, the more appropriateness. | | |
| Usability | Appropriateness recognizability | MQR05 | UAp-1 - Description completeness | What proportion of usage scenarios is described in the metamodel specifications? | The closer to 1, the more complete. | | |
| | | MQR06 | UAp-2 - Demonstration coverage | What proportion of metamodel concepts requiring demonstration have demonstration capability? | The closer to 1, the more capable. | | |
| | | MQR07 | UAp-3 - Evident concepts | What proportion of metamodel concepts are evident to the user? | The closer to 1, the better. | | |
| | | MQR08 | UAp-4 - Concept understandability | What proportion of metamodel concepts are correctly understood without prior training? | The closer to 1, the better. | | |
| | Learnability | MQR09 | ULe-1 - User guide completeness | What proportion of metamodel concepts is described in the user documentation that enable the use of the metamodel? | The closer to 1, the more complete. | | |
| Maintainability | Modularity | MQR10 | MMo-1 - Coupling of concepts | How strongly are the concepts independent and how many concepts are free of impacts from changes to other metamodel concepts? | The closer to 1, the less coupling. | | |

| Characteristic | Sub-characteristic | MQR | Measures | Measure Description | Interpretation of the measurement value | Target value | Acceptable tolerance value |
|---|---|---|---|---|---|---|---|
| | | MQR11 | MMo-2 - Complexity of exercise | How complex is building terminal models by analyzing the structure of the metamodel? | The higher, the more complex, i.e., the metamodel requires more ordered actions when creating the model elements. | | |
| | Reusability | MQR12 | MRe-1 - Reusability per application domain | How reusable is the metamodel to an application domain? | The closer to 1, the better. | | |
| | Modifiability | MQR13 | MMd-1 - Conceptual stability | How stable is the metamodel specification during the metamodel's development life cycle? | The closer to 1, the more stable. | | |
| | | MQR14 | MMd-2 - Change recordability | Are changes to metamodel specifications recorded adequately? | The closer to 1, the more recordable. The change control 0 indicates poor change control. | | |
| | | MQR15 | MMd-3 - Change impact | What is the frequency of adverse impacts after modification? | The closer to 1, the better. | | |
| | | | MMd-4 - Modification impact localization | How large is the impact of the modification on the metamodel? | The closer to 0, the lesser impact of modification. | | |
| | | | MMd-5 - Modification correcteness | What proportion of modifications has been implemented correctly? | The closer to 1, the better. | | |
| Portability | Adaptability | MQR16 | PAd-1 - Adaptability per application domain | How adaptable is the metamodel to an application domain? | The closer to 1, the better | | |
| | Replaceability | MQR17 | PRe-1 - Usage similarity | What proportion of usage scenarios of the replaced metamodel can be modeled without any additional learning or workaround? | The closer to 1, the better. | | |
| | | MQR18 | PRe-2 - Metamodel quality equivalence | What proportion of the quality measures is satisfied after replacing previous metamodel by this one? | The closer to 1, the better. | | |
| | | MQR19 | PRe-3 - Conceptual inclusiveness | Can the similar concepts easily be used after replacing previous metamodel by this one? | The closer to 1, the better. | | |

# ANNEX G - CRITERIA FOR EVALUATING THE METAMODEL

| Characteristic | Sub-characteristic | Quality Requirements | Measures | Sub-characteristic formula | Characteristic formula |
|---|---|---|---|---|---|
| Compliance | Conceptual compliance | MQR01 - The metamodel conceptual foundation must comply with widely-accepted and sound theories, regulations, standards, and conventions. | CCc-1 - Conceptual foundation | | |
| | | | CCc-2 - Backward Traceability | | |
| Conceptual suitability | Conceptual completeness | MQR02 - The metamodel must cover the concepts found in its specifications. | CCp-1 - Conceptual coverage | | |
| | Conceptual correctness | MQR03 - The metamodel must represent the concepts found in its specifications correctly. | CCr-1 - Conceptual correctness | | |
| | Conceptual appropriateness | MQR04 - The metamodel must represent the concepts required for achieving specific usage objectives. | CAp-1 - Conceptual appropriateness of usage objective | | |
| | | | CAp-2 - Conceptual appropriateness of metamodel | | |
| Usability | Appropriateness recognizability | MQR05 - The users must be able to recognize whether a metamodel is appropriate for their needs accordingly the usage scenarios described in the user documents. | UAp-1 - Description completeness | | |
| | | MQR06 - The users must be able to recognize whether a metamodel is appropriate for their needs accordingly the demonstration features of metamodel concepts. | UAp-2 - Demonstration coverage | | |
| | | MQR07 - The users must be able to recognize whether a metamodel is appropriate for their needs accordingly the evident concepts to the user in the metamodel specifications. | UAp-3 - Evident concepts | | |
| | | MQR08 - The users must be able to recognize whether a metamodel contain concepts whose purpose is correctly understood without prior training. | UAp-4 - Concept understandability | | |
| | Learnability | MQR09 - The users must be able to recognize whether a metamodel is appropriate for their needs accordingly the metamodel user documentation. | ULe-1 - User guide completeness | | |

| Characteristic | Sub-characteristic | Quality Requirements | Measures | Sub-characteristic formula | Characteristic formula |
|---|---|---|---|---|---|
| Maintainability | Modularity | MQR10 - The metamodel must be composed of discrete concepts such that a change of one concept has minimal impact on other concepts. | MMo-1 - Coupling of concepts | | |
| | | MQR11 - The metamodel must be composed of discrete concepts such that a creation of model elements does not enforce ordered modelling actions. | MMo-2 - Complexity of exercise | | |
| | Reusability | MQR12 - The metamodel must be able to be reused to modelling usage scenarios for different application domains. | MRe-1 - Reusability per application domain | | |
| | Modifiability | MQR13 - The users must be able to recognize metamodel modifications acoordingly the changes documented in the metamodel specification during metamodel development life cycle. | MMd-1 - Conceptual stability | | |
| | | MQR14 - The users must be able to recognize metamodel modifications acoordingly the change comments confirmed in review. | MMd-2 - Change recordability | | |
| | | MQR15 - The metamodel must be reused modified without introducing inconsistencies or degrading metamodel quality. | MMd-3 - Change impact | | |
| | | | MMd-4 - Modification impact localization | | |
| | | | MMd-5 - Modification correcteness | | |
| Portability | Adaptability | MQR16 - The metamodel must be able to be adapted to modelling usage scenarios for different application domains. | PAd-1 - Adaptability per application domain | | |
| | Replaceability | MQR17 - The metamodel must be able to replace another specified metamodel for the same purpose in the same application domain, without introducing any additional learning or workaround. | PRe-1 - Usage similarity | | |
| | | MQR18 - The metamodel must be able to replace another specified metamodel for the same purpose in the same application domain, without degrading metamodel quality degree. | PRe-2 - Metamodel quality equivalence | | |
| | | MQR19 - The metamodel must be able to replace another specified metamodel for the same purpose in the same application domain by using similar concepts of previous metamodel. | PRe-3 - Conceptual inclusiveness | | |

# ANNEX H – MEASUREMENTS TABLE

| Characteristic | Sub-characteristic | Quality Requirement | Measure | Measured value | Final measurement value | Sub-characteristic value | Characteristic value |
|---|---|---|---|---|---|---|---|
| Compliance | Conceptual compliance | MQR01 - The metamodel conceptual foundation must comply with widely-accepted and sound theories, regulations, standards, and conventions. | CCc-1 Conceptual foundation | | | | |
| | | | CCc-2 Backward Traceability | | | | |
| Conceptual suitability | Conceptual completeness | MQR02 - The metamodel must cover the concepts found in its specifications. | Ccp-1 Conceptual coverage | | | | |
| | Conceptual correctness | MQR03 - The metamodel must represent the concepts found in its specifications correctly. | CCr-1 Conceptual correctness | | | | |
| | Conceptual appropriateness | MQR04 - The metamodel must represent the concepts required for achieving specific usage objectives. | CAp-1 Conceptual appropriat. of usage objective | | | | |
| | | | CAp-2 Conceptual appropriat. of metamodel | | | | |
| Usability | Appropriateness recognizability | MQR05 - The users must be able to recognize whether a metamodel is appropriate for their needs accordingly the usage scenarios described in the user documents. | UAp-1 Description completeness | | | | |
| | | MQR06 - The users must be able to recognize whether a metamodel is appropriate for their needs accordingly the demonstration features of metamodel concepts. | UAp-2 Demonstration coverage | | | | |
| | | MQR07 - The users must be able to recognize whether a metamodel is appropriate for their needs accordingly the evident concepts to the user in the metamodel specifications. | UAp-3 Evident concepts | | | | |
| | | MQR08 - The users must be able to recognize whether a metamodel contain concepts whose purpose is correctly understood without prior training. | UAP-4 Concept understandability | | | | |
| | Learnability | MQR09 - The users must be able to recognize whether a metamodel is appropriate for their needs accordingly the metamodel user documentation. | ULe-1 User guide completeness | | | | |

| Characteristic | Sub-characteristic | Quality Requirement | Measure | Measured value | Final measurement value | Sub-characteristic value | Characteristic value |
|---|---|---|---|---|---|---|---|
| Maintainability | Modularity | MQR10 - The metamodel must be composed of discrete concepts such that a change of one concept has minimal impact on other concepts. | MMo-1 Coupling of concepts | | | | |
| | | MQR11 - The metamodel must be composed of discrete concepts such that a creation of model elements does not enforce ordered modelling actions. | MMo-2 Complexity of exercise | | | | |
| | Reusability | MQR12 - The metamodel must be able to be reused to modelling usage scenarios for different application domains. | MRe-1 Reusability per application domain | | | | |
| | Modifiability | MQR13 - The users must be able to recognize metamodel modifications acoordingly the changes documented in the metamodel specification during metamodel development life cycle. | MMd-1 Conceptual stability | | | | |
| | | MQR14 - The users must be able to recognize metamodel modifications acoordingly the change comments confirmed in review. | MMd-2 Change recordability | | | | |
| | | MQR15 - The metamodel must be reused modified without introducing inconsistencies or degrading metamodel quality. | MMd-3 Change impact | | | | |
| | | | MMd-4 Modific. impact localization | | | | |
| | | | MMd-5 Modification correctness | | | | |
| Portability | Adaptability | MQR16 - The metamodel must be able to be adapted to modelling usage scenarios for different application domains. | PAd-1 Adaptability per application domain | | | | |
| | Replaceability | MQR17 - The metamodel must be able to replace another specified metamodel for the same purpose in the same application domain, without introducing any additional learning or workaround. | PRe-1 Usage similarity | | | | |
| | | MQR18 - The metamodel must be able to replace another specified metamodel for the same purpose in the same application domain, without degrading metamodel quality degree. | PRe-2 Metamodel quality equivalence | | | | |
| | | MQR19 - The metamodel must be able to replace another specified metamodel for the same purpose in the same application domain by using similar concepts of previous metamodel. | PRe-3 Conceptual inclusiveness | | | | |

# ANNEX I - METAMODEL QUALITY EVALUATION PLAN (MQEP)
_________________________________________________________________

**Metamodel identification:** _______________________________________________

Evaluation Requester: _________________________________ Plan elaboration date:  /    /

## 1. Evaluation Requirements

### 1.1. Purpose

Choose the evaluation purpose according to the metamodel version available for quality evaluation:

| Ongoing metamodel version | Select |
|---|---|
| Assure quality for the metamodel | |
| Decide on the acceptance of an intermediate metamodel version | |
| Access the ongoing feasibility of the ongoing metamodel | |
| Predict or estimate final metamodel quality | |
| Discover improvement points in the metamodel | |
| Collect information on intermediate metamodel version in order to control and manage the process | |
| **Final metamodel version** | **Select** |
| Decide on the acceptance of the metamodel | |
| Compare a metamodel with others | |
| Select a metamodel from among alternative metamodels | |
| Assess both positive and negative effects of a metamodel | |
| Discover improvement points in the metamodel | |

### 1.2. Metamodel artifacts

The availability of metamodel-related artifacts is crucial to enable the evaluation of metamodel quality requirements. Choose the artifacts that will be available during the quality evaluation process:

| Metamodel artifacts available | Select |
|---|---|
| Metamodel specifications (requirements and design documents) | |
| Metamodel implementation | |
| Metamodel user documentation | |
| Metamodel history documentation | |
| Specification of different application domains | |
| Metamodel to be replaced by the evaluated metamodel | |

### 1.3. Resources

Evaluation resources must be defined, such as personnel, roles involved in the evaluation, evaluation tools to be used, evaluation budget, evaluation deadlines, and examples of metamodel application domains.

## 2. Metamodel Quality Requirements

The availability of specific metamodel artifacts is required to evaluate each metamodel quality requirement (MQR). Then, select each MQR for evaluation according to the available artifacts for you:

| Quality Requirements | Metamodel artifacts required | Select |
|---|---|---|
| MQR01 - The metamodel conceptual foundation must comply with widely-accepted and sound theories, regulations, standards, and conventions. | - Metamodel specifications | |
| MQR02 - The metamodel must cover the concepts found in its specifications. | - Metamodel specifications<br>- Metamodel implementation | |
| MQR03 - The metamodel must represent the concepts found in its specifications correctly. | - Metamodel specifications<br>- Metamodel implementation | |
| MQR04 - The metamodel must represent the concepts required for achieving specific usage objectives. | - Metamodel specifications<br>- Metamodel implementation<br>- Metamodel user documentation | |
| MQR05 - The users must be able to recognize whether a metamodel is appropriate for their needs accordingly the usage scenarios described in the user documents. | - Metamodel specifications<br>- Metamodel user documentation | |
| MQR06 - The users must be able to recognize whether a metamodel is appropriate for their needs accordingly the demonstration features of metamodel concepts. | - Metamodel specifications<br>- Metamodel user documentation | |
| MQR07 - The users must be able to recognize whether a metamodel is appropriate for their needs accordingly the evident concepts to the user in the metamodel specifications. | - Metamodel specifications<br>- Metamodel implementation | |
| MQR08 - The users must be able to recognize whether a metamodel contain concepts whose purpose is correctly understood without prior training. | - Metamodel specifications<br>- Metamodel implementation | |
| MQR09 - The users must be able to recognize whether a metamodel is appropriate for their needs accordingly the metamodel user documentation. | - Metamodel specifications<br>- Metamodel user documentation | |
| MQR10 - The metamodel must be composed of discrete concepts such that a change of one concept has minimal impact on other concepts. | - Metamodel implementation | |
| MQR11 - The metamodel must be composed of discrete concepts such that a creation of model elements does not enforce ordered modelling actions. | - Metamodel implementation | |
| MQR12 - The metamodel must be able to be reused to modelling usage scenarios for different application domains. | - Metamodel specifications<br>- Metamodel user documentation<br>- Specification of different application domains | |
| MQR13 - The users must be able to recognize metamodel modifications acoordingly the changes documented in the metamodel specification during metamodel development life cycle. | - Metamodel specifications<br>- Metamodel history documentation | |
| MQR14 - The users must be able to recognize metamodel modifications acoordingly the change comments confirmed in review. | - Metamodel specifications<br>- Metamodel history documentation | |
| MQR15 - The metamodel must be reused modified without introducing inconsistencies or degrading metamodel quality. | - Metamodel specifications<br>- Metamodel history documentation | |
| MQR16 - The metamodel must be able to be adapted to modelling usage scenarios for different application domains. | - Metamodel user documentation<br>- Specification of different application domains | |
| MQR17 - The metamodel must be able to replace another specified metamodel for the same purpose in the same application domain, without introducing any additional learning or workaround. | - Metamodel user documentation<br>- Metamodel to be replaced by the evaluated metamodel | |
| MQR18 - The metamodel must be able to replace another specified metamodel for the same purpose in the same application domain, without degrading metamodel quality degree. | - Metamodel specifications<br>- Metamodel implementation<br>- Metamodel to be replaced by the evaluated metamodel | |
| MQR19 - The metamodel must be able to replace another specified metamodel for the same purpose in the same application domain by using similar concepts of previous metamodel. | - Metamodel specifications<br>- Metamodel implementation | |

## 3. Metamodel Quality Measures

The selected quality measures must cover all chosen MQRs. Observe to which quality sub-characteristic (and characteristic) each measure is associated. Now, select quality measures according to the MQRs chosen for this evaluation:

| Characteristic | Sub-characteristic | MQR | Measures | Measure Description | Measurement function | Select |
|---|---|---|---|---|---|---|
| Compliance | Conceptual compliance | MQR01 | CCc-1 - Conceptual foundation | Which widely-accepted and sound theories, regulations, standards, and conventions is the metamodel compliant to? | A nominal list of widely-accepted and sound theories, regulations, standards, and conventions to which the metamodel is compliant. | |
| | | | CCc-2 - Backward Traceability | Which are the metamodel concepts that can be traced back to their conceptual foundations? | A nominal list of each metamodel concept with its respective conceptual foundation. | |
| Conceptual suitability | Conceptual completeness | MQR02 | CCp-1 - Conceptual coverage | What proportion of the specified concepts has been modeled? | $X = 1 - A / B$<br>A = Number of missing concepts.<br>B = Number of concepts described in the metamodel specification.<br>$0 <= X <= 1$.<br>The closer to 1, the more complete. | |
| | Conceptual correctness | MQR03 | CCr-1 - Conceptual correctness | What proportion of metamodel concepts are modeled correctly? | $X = 1 - A / B$<br>A = Number of incorrectly modeled concepts.<br>B = Number of concepts considered in the evaluation.<br>$0 <= X <= 1$.<br>The closer to 1, the more correct. | |
| | Conceptual appropriateness | MQR04 | CAp-1 - Conceptual appropriateness of usage objective | What proportion of the metamodel concepts provides appropriate outcome to achieve a specific usage objective? | $X = 1 - A / B$<br>A = Number of missing or incorrectly modeled concepts among those that are required for achieving a specific usage objective.<br>B = Number of concepts required for achieving a specific usage objective.<br>$0 <= X <= 1$.<br>The closer to 1, the more appropriateness. | |
| | | | CAp-2 - Conceptual appropriateness of metamodel | What proportion of the metamodel concepts required by the users to achieve their objectives provides appropriate outcome? | $$x = \sum_{i=1 \text{ to } n} A_i / n$$<br>Ai = Appropriateness score for usage objective i, that is, the measured value of CAp-1 for i-th specific usage objective.<br>N = Number of usage objectives.<br>The closer to 1, the more appropriateness. | |

| Characteristic | Sub-characteristic | MQR | Measures | Measure Description | Measurement function | Select |
|---|---|---|---|---|---|---|
| Usability | Appropriateness recognizability | MQR05 | UAp-1 - Description completeness | What proportion of usage scenarios is described in the metamodel specifications? | X = A / B<br>A = Number of usage scenarios described in the user documents that match usage scenarios described in the metamodel specifications.<br>B = Number of usage scenarios described in the metamodel specifications.<br>0 <= X <= 1.<br>The closer to 1, the more complete. | |
| | | MQR06 | UAp-2 - Demonstration coverage | What proportion of metamodel concepts requiring demonstration have demonstration capability? | X = A / B<br>A = Number of concepts with demonstration features.<br>B = Number of concepts that could benefit from demonstration features.<br>0 <= X <= 1.<br>The closer to 1, the more capable. | |
| | | MQR07 | UAp-3 - Evident concepts | What proportion of metamodel concepts are evident to the user? | X = A/B<br>A = Number of concepts evident to the user.<br>B = Number of concepts described in the metamodel specification.<br>0 <= X <= 1.<br>The closer to 1, the better. | |
| | | MQR08 | UAp-4 - Concept understandability | What proportion of metamodel concepts are correctly understood without prior training? | X = A / B<br>A = Number of concepts whose purpose is correctly understood without prior training.<br>B = Number of concepts described in the metamodel specification.<br>0 <= X <= 1.<br>The closer to 1, the better. | |
| | Learnability | MQR09 | ULe-1 - User guide completeness | What proportion of metamodel concepts is described in the user documentation that enable the use of the metamodel? | X = A / B<br>A = Number of concepts described in the user documentation as required.<br>B = Number of concepts required to be documented.<br>0 <= X <= 1.<br>The closer to 1, the more complete. | |

| Characteristic | Sub-characteristic | MQR | Measures | Measure Description | Measurement function | Select |
|---|---|---|---|---|---|---|
| Maintainability | Modularity | MQR10 | MMo-1 - Coupling of concepts | How strongly are the concepts independent and how many concepts are free of impacts from changes to other metamodel concepts? | X = A / B<br>A = Number of concepts with no impact on others<br>B = Number of specified concepts which are required to be independent.<br>0 <= X <= 1.<br>The closer to 1, the less coupling. | |
| | | MQR11 | MMo-2 - Complexity of exercise | How complex is building terminal models by analyzing the structure of the metamodel? | X = A − B<br>A = Number of instantiation elements that must be done in order<br>B = Number of instantiation groups that must be completed, but in any order<br>The higher, the more complex, i.e., the metamodel requires more ordered actions when creating the model elements. | |
| | Reusability | MQR12 | MRe-1 - Reusability per application domain | How reusable is the metamodel to an application domain? | X = 1 - A / B<br>A = Number of usage scenarios which were not possible to be reused for an application domain in particular<br>B = Number of usage scenarios described in the metamodel specifications<br>0 <= X <= 1.<br>The closer to 1, the better. | |
| | Modifiability | MQR13 | MMd-1 - Conceptual stability | How stable is the metamodel specification during the metamodel's development life cycle? | X = 1 - A / B<br>A = Number of concepts changed during the metamodel's development life cycle.<br>B = Number of concepts described in the metamodel specification.<br>0 <= X <= 1<br>The closer to 1, the more stable. | |
| | | MQR14 | MMd-2 - Change recordability | Are changes to metamodel specifications recorded adequately? | X = A / B<br>A = Number of changes in concepts having change comments confirmed in review.<br>B = Number of concepts changed from original metamodel specification.<br>0 <= X <= 1.<br>The closer to 1, the more recordable.<br>The change control 0 indicates poor change control. | |

| Characteristic | Sub-characteristic | MQR | Measures | Measure Description | Measurement function | Select |
|---|---|---|---|---|---|---|
| | | MQR15 | MMd-3 - Change impact | What is the frequency of adverse impacts after modification? | X = 1 - A / B<br>A = Number of detected adverse impacts after modifications.<br>B = Number of modifications made.<br>0 <= X <= 1<br>The closer to 1, the better. | |
| | | | MMd-4 - Modification impact localization | How large is the impact of the modification on the metamodel? | X = A / B<br>A = Number of concepts affected by modification, confirmed in review.<br>B = Number of concepts described in the metamodel specification.<br>0 <= X <= 1.<br>The closer to 0, the lesser impact of modification. | |
| | | | MMd-5 - Modification correctness | What proportion of modifications has been implemented correctly? | X = 1 - A / B<br>A = Number of modifications that caused an adverse impact within a defined period after made.<br>B = Number of modifications made.<br>0 <= X <= 1<br>The closer to 1, the better. | |
| Portability | Adaptability | MQR16 | PAd-1 - Adaptability per application domain | How adaptable is the metamodel to an application domain? | X = 1 - A / B<br>A = Number of usage scenarios which were not possible to be modeled for an application domain in particular<br>B = Number of usage scenarios described in the metamodel specifications<br>0 <= X <= 1<br>The closer to 1, the better | |
| | Replaceability | MQR17 | PRe-1 - Usage similarity | What proportion of usage scenarios of the replaced metamodel can be modeled without any additional learning or workaround? | X = A / B<br>A = Number of usage scenarios which can be modeled without any additional learning or workaround<br>B = Number of usage scenarios in the replaced metamodel<br>0 <= X <= 1.<br>The closer to 1, the better. | |

| Characteristic | Sub-characteristic | MQR | Measures | Measure Description | Measurement function | Select |
|---|---|---|---|---|---|---|
| | | MQR18 | PRe-2 - Metamodel quality equivalence | What proportion of the quality measures is satisfied after replacing previous metamodel by this one? | X = A / B<br>A = Number of quality measures of the new metamodel which are better or equal to the replaced metamodel<br>B = Number of quality measures of the replaced metamodel that are relevant<br>0 <= X <= 1.<br>The closer to 1, the better. | |
| | | MQR19 | PRe-3 - Conceptual inclusiveness | Can the similar concepts easily be used after replacing previous metamodel by this one? | X = A / B<br>A = Number of concepts which produce similar results as before<br>B = Number of concepts which have to be used in the replaced metamodel<br>0 <= X <= 1.<br>The closer to 1, the better. | |

## 4. Criteria for Metamodel Quality Measures

Measures decision criteria are numerical thresholds or targets used to determine some needs, or describe the level of confidence in a given result. Next, define a target value and an acceptable tolerance value to each chosen measure:

| Characteristic | Sub-characteristic | MQR | Measures | Measure Description | Interpretation of the measurement value | Target value | Acceptable tolerance value |
|---|---|---|---|---|---|---|---|
| Compliance | Conceptual compliance | MQR01 | CCc-1 - Conceptual foundation | Which widely-accepted and sound theories, regulations, standards, and conventions is the metamodel compliant to? | | | |
| | | | CCc-2 - Backward Traceability | Which are the metamodel concepts that can be traced back to their conceptual foundations? | | | |
| Conceptual suitability | Conceptual completeness | MQR02 | CCp-1 - Conceptual coverage | What proportion of the specified concepts has been modeled? | The closer to 1, the more complete. | | |
| | Conceptual correctness | MQR03 | CCr-1 - Conceptual correctness | What proportion of metamodel concepts are modeled correctly? | The closer to 1, the more correct. | | |
| | Conceptual appropriateness | MQR04 | CAp-1 - Conceptual appropriateness of usage objective | What proportion of the metamodel concepts provides appropriate outcome to achieve a specific usage objective? | The closer to 1, the more appropriateness. | | |
| | | | CAp-2 - Conceptual appropriateness of metamodel | What proportion of the metamodel concepts required by the users to achieve their objectives provides appropriate outcome? | The closer to 1, the more appropriateness. | | |
| Usability | Appropriateness recognizability | MQR05 | UAp-1 - Description completeness | What proportion of usage scenarios is described in the metamodel specifications? | The closer to 1, the more complete. | | |
| | | MQR06 | UAp-2 - Demonstration coverage | What proportion of metamodel concepts requiring demonstration have demonstration capability? | The closer to 1, the more capable. | | |
| | | MQR07 | UAp-3 - Evident concepts | What proportion of metamodel concepts are evident to the user? | The closer to 1, the better. | | |
| | | MQR08 | UAp-4 - Concept understandability | What proportion of metamodel concepts are correctly understood without prior training? | The closer to 1, the better. | | |
| | Learnability | MQR09 | ULe-1 - User guide completeness | What proportion of metamodel concepts is described in the user documentation that enable the use of the metamodel? | The closer to 1, the more complete. | | |
| Maintainability | Modularity | MQR10 | MMo-1 - Coupling of concepts | How strongly are the concepts independent and how many concepts are free of impacts from changes to other | The closer to 1, the less coupling. | | |

| Characteristic | Sub-characteristic | MQR | Measures | Measure Description | Interpretation of the measurement value | Target value | Acceptable tolerance value |
|---|---|---|---|---|---|---|---|
| | | | | metamodel concepts? | | | |
| | | MQR11 | MMo-2 - Complexity of exercise | How complex is building terminal models by analyzing the structure of the metamodel? | The higher, the more complex, i.e., the metamodel requires more ordered actions when creating the model elements. | | |
| | Reusability | MQR12 | MRe-1 - Reusability per application domain | How reusable is the metamodel to an application domain? | The closer to 1, the better. | | |
| | Modifiability | MQR13 | MMd-1 - Conceptual stability | How stable is the metamodel specification during the metamodel's development life cycle? | The closer to 1, the more stable. | | |
| | | MQR14 | MMd-2 - Change recordability | Are changes to metamodel specifications recorded adequately? | The closer to 1, the more recordable. The change control 0 indicates poor change control. | | |
| | | MQR15 | MMd-3 - Change impact | What is the frequency of adverse impacts after modification? | The closer to 1, the better. | | |
| | | | MMd-4 - Modification impact localization | How large is the impact of the modification on the metamodel? | The closer to 0, the lesser impact of modification. | | |
| | | | MMd-5 - Modification correcteness | What proportion of modifications has been implemented correctly? | The closer to 1, the better. | | |
| Portability | Adaptability | MQR16 | PAd-1 - Adaptability per application domain | How adaptable is the metamodel to an application domain? | The closer to 1, the better | | |
| | Replaceability | MQR17 | PRe-1 - Usage similarity | What proportion of usage scenarios of the replaced metamodel can be modeled without any additional learning or workaround? | The closer to 1, the better. | | |
| | | MQR18 | PRe-2 - Metamodel quality equivalence | What proportion of the quality measures is satisfied after replacing previous metamodel by this one? | The closer to 1, the better. | | |
| | | MQR19 | PRe-3 - Conceptual inclusiveness | Can the similar concepts easily be used after replacing previous metamodel by this one? | The closer to 1, the better. | | |

## 5. Criteria for Evaluating the Metamodel

Evaluation decision criteria shall be defined with separate evaluation criteria for different quality characteristics, which may be in terms of individual quality sub-characteristics and measures. Now, register the formulas that will be used to calculate characteristics' and sub-characteristics' grades:

| Characteristic | Sub-characteristic | Quality Requirements | Measures | Sub-characteristic formula | Characteristic formula |
|---|---|---|---|---|---|
| Compliance | Conceptual compliance | MQR01 - The metamodel conceptual foundation must comply with widely-accepted and sound theories, regulations, standards, and conventions. | CCc-1 - Conceptual foundation | | |
| | | | CCc-2 - Backward Traceability | | |
| Conceptual suitability | Conceptual completeness | MQR02 - The metamodel must cover the concepts found in its specifications. | CCp-1 - Conceptual coverage | | |
| | Conceptual correctness | MQR03 - The metamodel must represent the concepts found in its specifications correctly. | CCr-1 - Conceptual correctness | | |
| | Conceptual appropriateness | MQR04 - The metamodel must represent the concepts required for achieving specific usage objectives. | CAp-1 - Conceptual appropriateness of usage objective | | |
| | | | CAp-2 - Conceptual appropriateness of metamodel | | |
| Usability | Appropriateness recognizability | MQR05 - The users must be able to recognize whether a metamodel is appropriate for their needs accordingly the usage scenarios described in the user documents. | UAp-1 - Description completeness | | |
| | | MQR06 - The users must be able to recognize whether a metamodel is appropriate for their needs accordingly the demonstration features of metamodel concepts. | UAp-2 - Demonstration coverage | | |
| | | MQR07 - The users must be able to recognize whether a metamodel is appropriate for their needs accordingly the evident concepts to the user in the metamodel specifications. | UAp-3 - Evident concepts | | |
| | | MQR08 - The users must be able to recognize whether a metamodel contain concepts whose purpose is correctly understood without prior training. | UAp-4 - Concept understandability | | |
| | Learnability | MQR09 - The users must be able to recognize whether a metamodel is appropriate for their needs accordingly the metamodel user documentation. | ULe-1 - User guide completeness | | |

| Characteristic | Sub-characteristic | Quality Requirements | Measures | Sub-characteristic formula | Characteristic formula |
|---|---|---|---|---|---|
| Maintainability | Modularity | MQR10 - The metamodel must be composed of discrete concepts such that a change of one concept has minimal impact on other concepts. | MMo-1 - Coupling of concepts | | |
| | | MQR11 - The metamodel must be composed of discrete concepts such that a creation of model elements does not enforce ordered modelling actions. | MMo-2 - Complexity of exercise | | |
| | Reusability | MQR12 - The metamodel must be able to be reused to modelling usage scenarios for different application domains. | MRe-1 - Reusability per application domain | | |
| | Modifiability | MQR13 - The users must be able to recognize metamodel modifications acoordingly the changes documented in the metamodel specification during metamodel development life cycle. | MMd-1 - Conceptual stability | | |
| | | MQR14 - The users must be able to recognize metamodel modifications acoordingly the change comments confirmed in review. | MMd-2 - Change recordability | | |
| | | MQR15 - The metamodel must be reused modified without introducing inconsistencies or degrading metamodel quality. | MMd-3 - Change impact | | |
| | | | MMd-4 - Modification impact localization | | |
| | | | MMd-5 - Modification correcteness | | |
| Portability | Adaptability | MQR16 - The metamodel must be able to be adapted to modelling usage scenarios for different application domains. | PAd-1 - Adaptability per application domain | | |
| | Replaceability | MQR17 - The metamodel must be able to replace another specified metamodel for the same purpose in the same application domain, without introducing any additional learning or workaround. | PRe-1 - Usage similarity | | |
| | | MQR18 - The metamodel must be able to replace another specified metamodel for the same purpose in the same application domain, without degrading metamodel quality degree. | PRe-2 - Metamodel quality equivalence | | |
| | | MQR19 - The metamodel must be able to replace another specified metamodel for the same purpose in the same application domain by using similar concepts of previous metamodel. | PRe-3 - Conceptual inclusiveness | | |

## 6. Metamodel Evaluation Activities

Evaluation activities must be included in the evaluation schedule, containing the responsibilities of the parties involved in the evaluation and the activity deadlines. The schedule of activities will depend on the number of measures to be evaluated, the number of evaluators involved, the deadlines, among others items.

| Evaluation process activity | Evaluators | Start date | End date |
|---|---|---|---|
| **Execute the metamodel evaluation** | | | |
| *Make metamodel quality measurements* | | | |
|     CCc-1 Conceptual foundation | | | |
|     CCc-2 Backward Traceability | | | |
|     CCp-1 Conceptual coverage | | | |
|     CCr-1 Conceptual correctness | | | |
|     CAp-1 Conceptual appropriat. of usage objective | | | |
|     CAp-2 Conceptual appropriat. of metamodel | | | |
|     UAp-1 Description completeness | | | |
|     UAp-2 Demonstration coverage | | | |
|     UAp-3 Evident concepts | | | |
|     UAp-4 Concept understandability | | | |
|     ULe-1 User guide completeness | | | |
|     MMo-1 Coupling of concepts | | | |
|     MMo-2 Complexity of exercise | | | |
|     MRe-1 Reusability per application domain | | | |
|     MMd-1 Conceptual stability | | | |
|     MMd-2 Change recordability | | | |
|     MMd-3 Change impact | | | |
|     MMd-4 Modification impact localization | | | |
|     MMd-5 Modification correctness | | | |
|     PAd-1 Adaptability per application domain | | | |
|     PRe-1 Usage similarity | | | |
|     PRe-2 Metamodel quality equivalence | | | |
|     PRe-3 Conceptual inclusiveness | | | |
| *Apply decision criteria for quality measures* | | | |
| *Apply decision criteria for metamodel evaluation* | | | |
| **Conclude the metamodel evaluation** | | | |
| *Review the metamodel evaluation results* | | | |
| *Create the metamodel evaluation report* | | | |

## 7. Measurements Table

The measurements table must be filled in during the assessment by each evaluator to be included in the final metamodel evaluation report.

| Characteristic | Sub-characteristic | Quality Requirement | Measure | Measured value | Final measurement value | Sub-characteristic value | Characteristic value |
|---|---|---|---|---|---|---|---|
| Compliance | Conceptual compliance | MQR01 - The metamodel conceptual foundation must comply with widely-accepted and sound theories, regulations, standards, and conventions. | CCc-1 Conceptual foundation | | | | |
| | | | CCc-2 Backward Traceability | | | | |
| Conceptual suitability | Conceptual completeness | MQR02 - The metamodel must cover the concepts found in its specifications. | Ccp-1 Conceptual coverage | | | | |
| | Conceptual correcteness | MQR03 - The metamodel must represent the concepts found in its specifications correctly. | CCr-1 Conceptual correcteness | | | | |
| | Conceptual appropriateness | MQR04 - The metamodel must represent the concepts required for achieving specific usage objectives. | CAp-1 Conceptual appropriat. of usage objective | | | | |
| | | | CAp-2 Conceptual appropriat. of metamodel | | | | |
| Usability | Appropriateness recognizability | MQR05 - The users must be able to recognize whether a metamodel is appropriate for their needs accordingly the usage scenarios described in the user documents. | UAp-1 Description completeness | | | | |
| | | MQR06 - The users must be able to recognize whether a metamodel is appropriate for their needs accordingly the demonstration features of metamodel concepts. | UAp-2 Demonstration coverage | | | | |
| | | MQR07 - The users must be able to recognize whether a metamodel is appropriate for their needs accordingly the evident concepts to the user in the metamodel specifications. | UAp-3 Evident concepts | | | | |
| | | MQR08 - The users must be able to recognize whether a metamodel contain concepts whose purpose is correctly understood without prior training. | UAP-4 Concept understandability | | | | |
| | Learnability | MQR09 - The users must be able to recognize whether a metamodel is appropriate for their needs accordingly the metamodel user documentation. | ULe-1 User guide completeness | | | | |
| Maintainability | Modularity | MQR10 - The metamodel must be composed of discrete concepts such that a change of one concept has minimal impact on other concepts. | MMo-1 Coupling of concepts | | | | |

| Characteristic | Sub-characteristic | Quality Requirement | Measure | Measured value | Final measurement value | Sub-characteristic value | Characteristic value |
|---|---|---|---|---|---|---|---|
| | | MQR11 - The metamodel must be composed of discrete concepts such that a creation of model elements does not enforce ordered modelling actions. | MMo-2 Complexity of exercise | | | | |
| | Reusability | MQR12 - The metamodel must be able to be reused to modelling usage scenarios for different application domains. | MRe-1 Reusability per application domain | | | | |
| | Modifiability | MQR13 - The users must be able to recognize metamodel modifications acoordingly the changes documented in the metamodel specification during metamodel development life cycle. | MMd-1 Conceptual stability | | | | |
| | | MQR14 - The users must be able to recognize metamodel modifications acoordingly the change comments confirmed in review. | MMd-2 Change recordability | | | | |
| | | MQR15 - The metamodel must be reused modified without introducing inconsistencies or degrading metamodel quality. | MMd-3 Change impact | | | | |
| | | | MMd-4 Modific. impact localization | | | | |
| | | | MMd-5 Modification correctness | | | | |
| Portability | Adaptability | MQR16 - The metamodel must be able to be adapted to modelling usage scenarios for different application domains. | PAd-1 Adaptability per application domain | | | | |
| | Replaceability | MQR17 - The metamodel must be able to replace another specified metamodel for the same purpose in the same application domain, without introducing any additional learning or workaround. | PRe-1 Usage similarity | | | | |
| | | MQR18 - The metamodel must be able to replace another specified metamodel for the same purpose in the same application domain, without degrading metamodel quality degree. | PRe-2 Metamodel quality equivalence | | | | |
| | | MQR19 - The metamodel must be able to replace another specified metamodel for the same purpose in the same application domain by using similar concepts of previous metamodel. | PRe-3 Conceptual inclusiveness | | | | |

# ANNEX J - MQUARE QUALITY EVALUATION REPORT TEMPLATE

**Metamodel identification:** ______________________________________________

Evaluator: _________________________________ Evaluation period: _______________________

## 1. Quality evaluation plan

Provide an overview of the quality evaluation plan.

## 2. The evaluators and their qualifications

Describe a brief overview of the evaluators' qualifications.

## 3. Problems or workarounds in adverse events

Describe problems and workarounds that may have occurred during the evaluation.

## 4. The results from the measurements and analyses performed

Provide the measurements table filled with the measurements values and also a detailed analysis of the results.

## 5. Result of the evaluation

Describe the results of the evaluation with the summarization of grades by characteristics and sub-characteristics.